\documentclass{jfm}
\usepackage{graphicx}
\usepackage{epstopdf, epsfig}
\usepackage{subcaption}
\usepackage{amsfonts,amssymb,amstext,amsmath}
\usepackage{booktabs}

\usepackage{xparse}
\usepackage[dvipsnames,table]{xcolor}     
\usepackage{units}
\newcommand{\avg}[1]{\left\langle #1 \right\rangle} 
\newcommand{\avgA}[1]{\left\langle #1 \right\rangle_{\!A}}
\newcommand{\vu}{\boldsymbol{u}}
\newcommand{\vx}{\boldsymbol{x}}
\newcommand{\grad}{\nabla}

\newcommand\Nu{\mbox{\textit{Nu}}}
\newcommand\Ra{\mbox{\textit{Ra}}}

\shorttitle{Energy budget of superstructures in RBC}
\shortauthor{G. Green, D.G. Vlaykov, J.P. Mellado, M. Wilczek}

\title{Resolved energy budget of superstructures in Rayleigh-B\'{e}nard convection}
\author{Gerrit Green\aff{1,2}  
   Dimitar G. Vlaykov\aff{1,3},
   Juan Pedro Mellado\aff{4}\footnote{Current address: Department of Physics, Aerospace Engineering Division, Universitat Polit\`{e}cnica de Catalunya, C. Jordi Girona 1-3, 08034, Barcelona, Spain }
  \and Michael Wilczek\aff{1,2}
  \corresp{\email{michael.wilczek@ds.mpg.de}}}

\affiliation{\aff{1} Max Planck Institute for Dynamics and Self-Organization (MPI DS),
	Am Fa\ss berg 17, 37077 G\"ottingen, Germany
\aff{2} Faculty of Physics, University of G\"{o}ttingen, Friedrich-Hund-Platz 1, 37077 G\"{o}ttingen, Germany
\aff{3} Astrophysics Group, College of Engineering, Mathematics and Physical Sciences, University of Exeter, EX4 4QL Exeter, UK
\aff{4} Max Planck Institute for Meteorology, Bundesstra\ss e 53, 20146 Hamburg, Germany
}

\usepackage{hyperref}
\hypersetup{
	colorlinks=true,
	linkcolor={MidnightBlue},
	citecolor={MidnightBlue},
	urlcolor={MidnightBlue},	
}

\begin{document}

\maketitle

\begin{abstract}
Turbulent superstructures, i.e.~large-scale flow structures in turbulent flows,
play a crucial role in many geo- and astrophysical settings. In turbulent Rayleigh-B\'{e}nard convection, for example, horizontally extended coherent large-scale convection rolls emerge. Currently, a detailed understanding of the interplay of small-scale turbulent fluctuations and large-scale coherent structures is missing. Here, we investigate the resolved kinetic energy and temperature variance budgets by applying a filtering approach to direct numerical simulations of Rayleigh-B\'{e}nard convection at high aspect ratio. In particular, we focus on the energy transfer rate between large-scale flow structures and small-scale fluctuations. We show that the small scales primarily act as a dissipation for the superstructures. However, we find that the height-dependent energy transfer rate has a complex structure with distinct bulk and boundary layer features. Additionally, we observe that the heat transfer between scales mainly occurs close to the thermal boundary layer. Our results clarify the interplay of superstructures and turbulent fluctuations and may help to guide the development of an effective description of large-scale flow features in terms of reduced-order models.
\end{abstract}


\section{Introduction}
Many turbulent flows in nature, for example in the atmosphere or in the interior of stars and planets, are driven by thermal gradients, which lead to convection. A characteristic feature of these flows is the coexistence of large-scale order and smaller-scale fluctuations. Prominent examples are cloud streets in the atmosphere \citep{atkinson96rog} or solar granulation \citep{nordlund09lrisp}. Currently, little is known about the interplay of small-scale fluctuations and large-scale order, but a detailed understanding is important for the development of reduced-order models, e.g.~in climate science, as well as in geo- and astrophysical settings. Better understanding the coexistence of this large-scale order and turbulence in convective flows is one motivation for the current work.

Rayleigh-B\'{e}nard convection (RBC), a confined flow between a heated bottom plate and a cooled top plate, is an idealized system to study convection and has been successfully employed to understand various phenomena such as pattern formation, spatio-temporal chaos \citep{bodenschatz00arofm,getling98} and turbulence \citep{lohse10arofm,chilla12epje}. Rayleigh-B\'enard convection is governed by three non-dimensional parameters, the Rayleigh number $\Ra$, characterizing the strength of the thermal driving, the Prandtl number $\Pran$, which is the ratio between kinematic viscosity and thermal diffusivity, and the aspect ratio $\Gamma$ of the system's width to its height. Above the onset of convection, at which the heat transfer changes from conduction to convection, a rich dynamics can be observed (see, e.g.~\citet{bodenschatz00arofm}). Close to onset, the flow organizes into regular convection rolls. As the Rayleigh number is increased, the flow becomes increasingly complex. At moderate Rayleigh numbers in high aspect ratio RBC, the dynamics of the convection rolls becomes chaotic, exhibiting spiral defect chaos (SDC) (see, e.g.~\citet{morris93prl} for an early study, or \citet{bodenschatz00arofm} and references therein for an overview). At much higher Rayleigh numbers, the flow becomes turbulent and features prominent smaller-scale flow structures such as thermal plumes \citep{siggia94arof,grossmann04pof,lohse10arofm,schumacher18pre}. 

\begin{figure}
	\begin{subfigure}{0.475\linewidth}
		\centering
		\includegraphics[width=.975\textwidth]{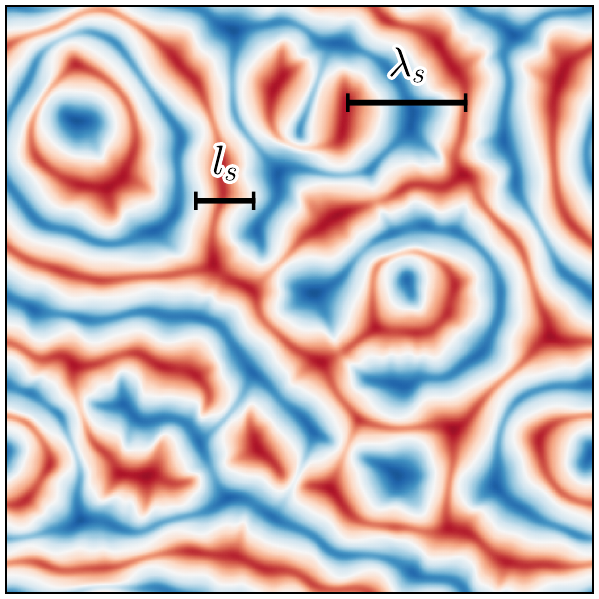}
		\caption{$\Ra=1.03\times10^4$ }
	\end{subfigure} 
	\begin{subfigure}{0.475\linewidth}
		\centering
		\includegraphics[width=.975\textwidth]{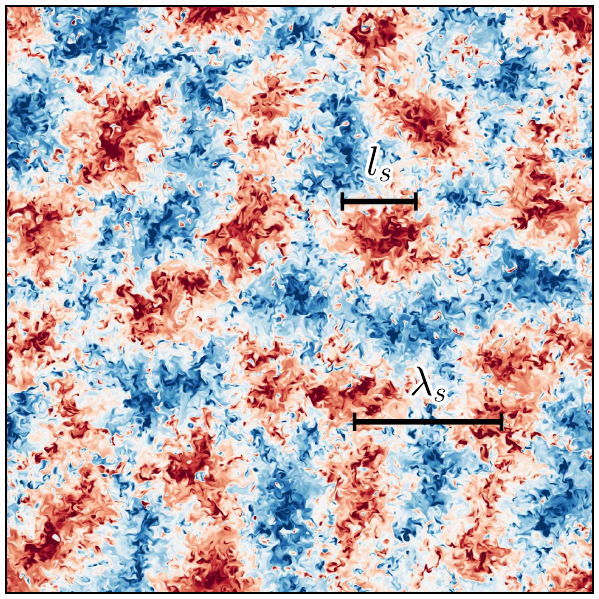} 
		\caption{$\Ra=1.07\times10^7$}
	\end{subfigure} 
	\caption{Temperature fields in the midplane for two different Rayleigh numbers with $\Pran=1$ and aspect ratio $24$. Red indicates hot rising fluid, and blue cold descending fluid. (a) Close to onset in the weakly nonlinear regime, regular patterns with wavelength $\lambda_s$ emerge. (b) Connected large-scale structures are present in the turbulent regime as well, and their length scale $\lambda_s$ is increased compared to onset. The small-scale fluctuations can be removed with a filter of width $l_s$, which preserves the large-scale rolls. For similar visualizations of turbulent superstructures, see also \citet{hartlep05jfm,stevens18prf,pandey18natcomm}.}
\label{fig:RBC_LSS}
\end{figure}

As visualized in figure \ref{fig:RBC_LSS}, even in the turbulent regime, horizontally extended large-scale convection rolls, so-called turbulent superstructures, have been observed in direct numerical simulations of large aspect ratio systems \citep{hartlep03prl,parodi04prl,shishkina06jfm,vonHardenberg08pra,emran15jfm,pandey18natcomm,stevens18prf,krug19arxiv}. Their large-scale structure and dynamics can be revealed, for example, by time averaging \citep{emran15jfm,pandey18natcomm}, and they are composed of clustered plumes \citep{parodi04prl}. The presence of the large-scale flow has important consequences for the temperature statistics in RBC, see \citet{lulff11njp,lulff15jfm,stevens18prf} as well as the heat transport \citep{stevens18prf,fonda19pnas}. Turbulent superstructures vary on time scales much larger than the characteristic free-fall time \citep{pandey18natcomm}, and their length scale increases with $\Ra$ \citep{hartlep03prl,hartlep05jfm,shishkina06jfm,pandey18natcomm,krug19arxiv}, which is visualized in figure \ref{fig:RBC_LSS}. 
Additionally, they appear to have a close connection to the boundary layer dynamics \citep{pandey18natcomm,stevens18prf}, e.g.~the local maxima and minima of the temperature in the midplane coincide with the position of hot and cold plume ridges in the boundary layer. 

For moderate Rayleigh numbers, the superstructure dynamics is reminiscent of SDC in the weakly nonlinear regime \citep{emran15jfm}. This points to the possibility of establishing connections to flows at much lower Rayleigh number, which are theoretically tractable by methods such as linear stability analysis and order parameter equations \citep{bodenschatz00arofm,manneville90}.
\captionsetup[subfigure]{singlelinecheck=off,justification=raggedright}
This is of considerable interest because, so far, only a few attempts exist to theoretically understand these turbulent large-scale patterns. \citet{elperin02pre,elperin06blm,elperin06pof} found large-scale instabilities based on a mean field theory combined with a turbulence closure. \citet{ibbeken19prl} studied the effect of small-scale fluctuations on large-scale patterns in a generalized Swift-Hohenberg model and showed that the fluctuations lead to an increased wavelength of the large-scale patterns. Still, the precise mechanism of the formation of the large-scale pattern and the selection of their length scale is not fully understood in turbulent RBC, and the emergence of large-scale rolls in the turbulent regime leaves many open questions. In particular, the interplay between superstructures and small-scale turbulence is currently largely unexplored. Thus, the main aim of this article is to clarify the impact of small-scale fluctuations and to characterize the energy budget of the large-scale convection rolls. With a focus on superstructures, this complements previous studies on the scale-resolved energy and temperature variance budgets of convective flows: \citet{togni15jfm} focused on the impact of thermal plumes and the scale dependence at different heights, \citet{kimmel00pof} and \citet{togni17pitvii,togni19jfm} aimed at improving large eddy simulations, \citet{valori20jfm} focused on small scales and \citet{faranda18jas} studied atmospheric flows.

Here, we investigate RBC by means of direct numerical simulations (DNS) in large aspect ratio systems from the weakly nonlinear regime close to onset up to the turbulent regime covering a Rayleigh number range from $\Ra=10^4$ to $\Ra=10^8$ at $\Pran=1$. To separate the scales, we apply a filtering approach \citep{germano92jfm} and isolate the superstructure dynamics. We then determine the energy and temperature variance budgets of the superstructures and the corresponding transfer rates between large-scale flow structures and small-scale fluctuations.
 
The remainder of the article is structured as follows. We first present the relevant theoretical and numerical background in section \ref{sec:methods}. In section \ref{sec:results}, the results are presented. Here, we find that at the scale of the superstructures the time- and volume-averaged resolved energy input into the large scales is primarily balanced by the energy transfer rate to small scales instead of the direct dissipation. To understand the role of the boundary layers, we supplement the volume-averaged analysis with a study of the height profiles of the different contributions to the resolved energy budget obtained from horizontal and time averages. We find that these profiles exhibit a complex near-wall structure and interpret the form of the profiles in terms of the plume dynamics. We complement the analysis of the resolved energy budget with that of the resolved temperature variance budget. This reveals that the averaged heat transfer rate is exceeded by the averaged direct thermal dissipation for all Rayleigh numbers, a qualitatively different behaviour than that of the energy transfer rate. Also, a substantial part of the heat transfer rate is limited to the boundary layers. Finally, we conclude in section \ref{sec:conclusions}.

\section{Theoretical and numerical background}
\label{sec:methods}
To begin with, we introduce the underlying equations and methods. We present the filtering approach as well as the resolved energy and temperature variance budgets used to study the transfer rates between scales. We then describe the numerical data used for our analysis.
\subsection{Governing equations}

RBC is governed by the Oberbeck-Boussinesq equations (OBEs), which describe the evolution of the velocity $\vu$ and the temperature fluctuation $\theta$, i.e.~the deviation from the mean temperature. In this set-up, it is assumed that the density varies linearly with temperature with only small variations, such that the fluid can still be considered as incompressible \citep{chilla12epje}.
Explicitly, the non-dimensionalized, three-dimensional equations are
\begin{subequations}\label{eq:OBE}
\begin{align}
\boldsymbol{\grad \cdot}\vu& =0 \label{eq:OBEdiv}\\
\partial_t \vu +\vu \boldsymbol{\cdot \grad} \vu & = - \boldsymbol{\grad} p^{\ast}+\sqrt{\frac{\Pran}{\Ra}} \grad^2\vu + \theta \hat{\boldsymbol{z}} \label{eq:OBEu}\\
\partial_t \theta+\vu \boldsymbol{\cdot \grad} \theta &= \frac{1}{\sqrt{\Ra\Pran}} \grad^2 \theta,\label{eq:OBEt}
\end{align}
\end{subequations}
in which $p^\ast$ is the kinematic pressure including gravity, which points in the negative $z$-direction. Here, $\hat{\boldsymbol{z}}$ is the unit vector in the vertical direction. 
The equations are non-dimensionalized with the temperature difference between top and bottom $\Delta$, the free-fall time $t_f=\sqrt{H/(\alpha g \Delta)}$ and the velocity $u_f=H/t_f$, where $H$ is the height of the system. The system is subject to two control parameters, the Prandtl number $\Pran=\nu/\kappa$, which is the ratio of kinematic viscosity to thermal diffusivity and the Rayleigh number $\Ra=g\alpha\Delta H^3/(\nu\kappa)$, the ratio between the strength of the thermal driving and damping by dissipation. Here, $g$ is the acceleration due to gravity and $\alpha$ the thermal expansion coefficient. These equations are supplemented with Dirichlet boundary conditions for the temperature as well as no-slip boundary conditions for the velocity at the top and bottom wall, and periodic boundary conditions at the side walls. 
Strong thermal driving leads to a turbulent convective flow at sufficiently high $\Ra$ far above the onset of convection.

In a statistically stationary state, exact relations between forcing and dissipation can be derived from the kinetic energy and temperature variance budgets \citep{shraiman90pra},
\begin{align}
\avg{\varepsilon}=\avg{ u_z\theta }&=\frac{1}{\sqrt{\Ra\Pran}}\left(\Nu-1\right), \label{eq:EkinV}\quad&\text{where}&~&\varepsilon&=\frac{1}{2}\sqrt{\frac{\Pran}{\Ra}}\left(\boldsymbol{\nabla}\boldsymbol{u}+\left(\boldsymbol{\nabla}\boldsymbol{u}\right)^\intercal\right)^2,\\ 
\text{and}\avg{ \chi }&=\frac{1}{\sqrt{\Ra\Pran}}\Nu \label{eq:EtempV},\quad
&\text{where}&~ &\chi&=\frac{1}{\sqrt{\Ra\Pran}}(\boldsymbol{\nabla}\theta)^2,
\end{align}
i.e.~the averaged energy input $\avg{ u_z\theta}$ is balanced by the averaged dissipation $\avg{\varepsilon}$, and the dimensionless heat transport $\Nu=\sqrt{\Ra\Pran}\avg{ u_z\theta}+1$ is balanced by the thermal dissipation $\avg{\chi }$. Here, $\avg{ \cdot }$ denotes an average over time and volume, which we simply refer to as volume averaged and ${}^\intercal$ stands for transpose. 
For more details, see also \citet{siggia94arof,chilla12epje} and \citet{ching14}. These statements for the averaged relation between forcing and dissipation are generalized to scale-dependent budgets in the following section.

\subsection{Filtering}

In order to separate small-scale fluctuations and large-scale structures, we use low-pass filtering. In this study, we only filter horizontally to extract the horizontally extended superstructures. 
Compared to three-dimensional filtering, this approach avoids complications in the interpretation of results introduced by the inhomogeneity in the vertical direction, especially near the boundaries \citep{sagaut06}. Note also that, besides a few exceptions, e.g.~\citet{fodor19blm}, this approach is widely used in the study of wall-bounded flows, see, e.g.~\citet{cimarelli11jfm,togni17pitvii,togni19jfm,bauer19prf} and \citet{valori20jfm}.
The filtering operator is a locally weighted average given by a convolution with a filter kernel $G_l$,
\begin{align}
\overline{\vu}_l(\vx)  = G_l \ast \vu = \frac{1}{l^2} \int_{x-l/2}^{x+l/2} \int_{y-l/2}^{y+l/2} \vu(x', y', z) \; \mathrm{d} x'\, \mathrm{d}y' \label{eq:filter}.
\end{align}
For our study, we choose a standard two-dimensional box filter. 
The large-scale velocity $\overline{\vu}_l$ encodes the velocity on scales larger than the scale $l$ in the horizontal directions.
The large-scale temperature $\overline{\theta}_l$ is defined analogously. In the following, we refer to scales below the filter width as unresolved and scales above it as resolved or large scale.
The evolution of the resolved scales is given by filtering \eqref{eq:OBE}
\begin{subequations}
\begin{align}
\boldsymbol{\grad \cdot} \overline{\vu}_l&=0 \label{eq:OBEdivl}\\
\partial_t \overline{\boldsymbol{u}}_l+\overline{\boldsymbol{u}}_l\boldsymbol{\cdot\nabla}\overline{\boldsymbol{u}}_l&=-\boldsymbol{\nabla} \overline{p}_l^{*}+ \sqrt{\frac{\Pran}{\Ra}}\nabla^2\overline{\boldsymbol{u}}_l+\overline{\theta}_l \hat{\boldsymbol{z}}-\boldsymbol{\nabla\cdot} \boldsymbol{\tau}_l \label{eq:OBEul}\\
\partial_t \overline{\theta}_l+\overline{\boldsymbol{u}}_l\boldsymbol{\cdot\nabla} \overline{\theta}_l&=\frac{1}{\sqrt{\Ra\Pran}}\nabla^2 \overline{\theta}_l-\boldsymbol{\nabla\cdot} {\boldsymbol{\gamma}}_l \label{eq:OBEtl},
\end{align}
\end{subequations}
in which 
\begin{align}
\boldsymbol{\tau}_l &= \overline{(\vu\vu)}_l-\overline{\vu}_l\overline{\vu}_l  \\
\text{and} \quad
\boldsymbol{\gamma}_l &=  \overline{(\vu\theta)}_l-\overline{\vu}_l\overline{\theta}_l \label{eq:heat_flux}.
\end{align}
Here additional terms involving $\boldsymbol{\tau}_l$ and $\boldsymbol{\gamma}_l$ appear due to the nonlinearity of the OBEs. The turbulent stress tensor $\boldsymbol{\tau}_l$ and turbulent heat flux $\boldsymbol{\gamma}_l$ effectively describe the impact of the unresolved scales on the resolved ones.

A few words on the limiting cases $l \to 0$ and $l \to \infty$ are in order. For any field $q$ 
\begin{align}
\lim_{l \to 0}	G_l \ast  q= q \label{eq:filter_lim0},
\end{align}
see, e.g.~\citet{sagaut06}.
On the other hand, for $l\to \infty$ the filtering is essentially a horizontal average, which we shall denote by $\avgA{\cdot}$, i.e.
\begin{align}
\lim_{l \to \infty}	G_l \ast q =\avgA{q} \label{eq:filter_liminf}.
\end{align}
This means that the filtering procedure applied in this work smoothly interpolates between the fully resolved and the height-dependent, horizontally averaged fields.
Using the above definitions, we derive the resolved energy budget in the next section. In particular, we focus on the resolved budgets at the scale of the turbulent superstructures.

\subsection{Resolved energy budget}
To derive the resolved energy budget, \eqref{eq:OBEul} is multiplied with $\overline{\vu}_l$, cf.~\citet{sagaut06,eyink95josp,eyink07coursenotes,eyink09pof,aluie09pof} and \citet{togni19jfm}. We obtain
\begin{align}
\partial_t e_l+\boldsymbol{\nabla\cdot}\boldsymbol{J}_l&=-\varepsilon_l
+Q_l-\Pi_l , \label{eq:resolved_budget}
\end{align}
and the individual terms are explicitly given by 
\begin{align}
   \varepsilon_l&=\frac{1}{2}\sqrt{\frac{\Pran}{\Ra}}\left(\boldsymbol{\nabla}\overline{\boldsymbol{u}}_l+\left(\boldsymbol{\nabla}\overline{\boldsymbol{u}}_l\right)^\intercal\right)^2,\\
 Q_l&= \overline{\theta}_l \overline{\boldsymbol{u}}_l\cdot\hat{\boldsymbol{z}}, \\
   \Pi_l&=- \left(\boldsymbol{\nabla}\overline{\boldsymbol{u}}_l\right):\boldsymbol{\tau}_l, \\
   \text{and} \quad
   \boldsymbol{J}_l&=\left(e_l+\overline{p}^*_l\right)\overline{\boldsymbol{u}}_l-\sqrt{\frac{\Pran}{\Ra}}\boldsymbol{\nabla} e_l+\boldsymbol{\tau}_l\cdot\overline{\boldsymbol{u}}_l-\sqrt{\frac{\Pran}{\Ra}}\overline{\boldsymbol{u}}_l\boldsymbol{\cdot \nabla}\overline{\boldsymbol{u}}_l  .
\end{align}
Here, $e_l=\overline{\vu}_l^2/2$ is the resolved kinetic energy, $\varepsilon_l$ denotes the direct large-scale dissipation and $Q_l$ is the energy input rate into the resolved scales by thermal driving. Compared to the unfiltered energy budget, an additional contribution $\Pi_l$ appears. It originates from the nonlinear term in the momentum equation and captures the transfer rate of kinetic energy between scales. It can act, depending on its sign, as a sink or source for the resolved scales. In the following, we refer to $\Pi_l$ as the energy transfer.
The evolution equation also contains a large-scale spatial flux term $\boldsymbol{J}_l$, which redistributes energy in space. As we focus on the energy transfer between scales in this study, we refrain from characterizing the individual contributions to the spatial flux. For a detailed study of the corresponding unfiltered spatial flux terms, we refer to \citet{petschel15jfm}.

In a nutshell, \eqref{eq:resolved_budget} describes the change of the resolved energy $e_l$ by spatial redistribution, direct dissipation, large-scale thermal driving and energy transfer between scales. Complementary to spectral analysis techniques (see, e.g.~\citet{domaradzki94pof,lohse10arofm,verma17njp} and \citet{verma18}), this approach allows the spatially resolved study of the energy transfer between superstructures and small-scale fluctuations. In the following, spatial and temporal averages of the resolved energy balance are considered.

\subsubsection{Averaged resolved energy budget}\label{sec:mean_energy_budget}
To derive a scale-resolved generalization of \eqref{eq:EkinV}, we average \eqref{eq:resolved_budget} over space and time. In a statistically stationary state, $\avg{ \partial_t e_l} $ vanishes. The averaged flux $\avg{ \nabla\cdot\boldsymbol{J}_l }$ vanishes as well because of the no-slip boundary conditions for the velocity. The resulting balance
\begin{align}
\avg{ Q_l}=\avg{\varepsilon_l}+\avg{\Pi_l } \label{eq:resolved_budget_global}
\end{align} 
shows that, at each scale, the energy input is balanced by the direct dissipation and the energy transfer between scales. Note that the latter is not present in the unfiltered energy balance \eqref{eq:EkinV}. As presented in Appendix \ref{app:Gammal}, \eqref{eq:resolved_budget_global} can also be related to the Nusselt number.

Because the energy dissipation primarily occurs at the smallest scales in three-dimensional turbulence \citep{pope2000}, the introduced energy has to be transferred to the dissipative scales for a statistically stationary state to exist. Since RBC is forced on all scales by buoyancy, including the largest scales, the volume-averaged energy transfer above the dissipative range is
\textit{a priori} expected to be down-scale. 
Accordingly, the volume-averaged energy transfer has to act as a sink in the resolved energy budget. 

To understand the scale dependence of the different contributions, we first determine the two limits $l \to 0$ and $l \to \infty$, for which we make use of \eqref{eq:filter_lim0} and \eqref{eq:filter_liminf}. For $l \to 0$, $\Pi_l$ vanishes and
\begin{align}
\lim_{l \to 0}\avg{ Q_l}-\avg{\varepsilon_l}-\avg{\Pi_l }=\avg{ Q}-\avg{\varepsilon}=0,
\end{align} 
i.e.~the unfiltered balance is recovered with $Q=u_z\theta$.
In the limit $l \to \infty$, the filtering is equivalent to a horizontal average. In an infinitely extended domain, $\avgA{\vu}=0$, and therefore, all terms in the budget vanish individually
\begin{align}
\lim_{l \to \infty}\avg{ Q_l}=\lim_{l \to \infty}\avg{\varepsilon_l}=\lim_{l \to \infty}\avg{\Pi_l }=0.
\end{align}

The detailed scale dependence and the balance between the different terms at the length scale corresponding to superstructures are investigated numerically and presented in subsequent sections.

To complete this section, we present the horizontally and time-averaged resolved kinetic energy budget
\begin{align}
\avgA{\nabla\cdot\boldsymbol{J}_l}&=-\avgA{ \varepsilon_l }-\avgA{ \Pi_l }
+\avgA{ Q_l} \label{eq:resolved_budget_profile},
\end{align}
in which $\avgA{\cdot }$ from now on describes a horizontal and time average.
This will be used to determine the role of the boundary layers and to refine the picture based on the volume average.
Compared to the volume-averaged resolved energy budget, the spatial flux term $\avgA{\nabla\cdot\boldsymbol{J}_l}$ does not vanish.
The limiting behaviour is very similar to that of the volume-averaged balance. As $l\to 0$, the energy transfer vanishes, whereas the other terms recover the unfiltered balance
\begin{align}
\avgA{\boldsymbol{\nabla\cdot}\boldsymbol{J}}&=-\avgA{ \varepsilon }
+\avgA{ Q}\label{eq:energy_budget},\\
\text{where} \quad \boldsymbol{J}&=\left(e+p^*\right)\boldsymbol{u}-\sqrt{\frac{\Pran}{\Ra}}\boldsymbol{\nabla} e-\sqrt{\frac{\Pran}{\Ra}}\boldsymbol{u\cdot \nabla}\boldsymbol{u} .
\end{align}
As $l\to \infty$, all terms vanish individually for the same reason as above.

In the work of \citet{petschel15jfm}, the unfiltered budget \eqref{eq:energy_budget} has been studied. It was shown that most of the energy is typically dissipated near the wall and energy input occurs in the bulk, from where it is transported to the wall. The generalization to a resolved energy budget allows us to investigate these processes as a function of scale, and in particular at the scale of the turbulent superstructures.

\subsection{Resolved temperature variance budget}
To complete the theoretical background, we consider the budget of the resolved temperature variance $e^\theta_l=\overline{\theta}^2_l/2$:
\begin{align}
\partial_t e^\theta_l+\boldsymbol{\nabla\cdot J}_l^\theta&=-\chi_l-\Pi_l^\theta,\label{eq:tempvar_l}
\end{align}
where the individual terms are given by
\begin{align}
\chi_l&=\frac{1}{\sqrt{\Ra\Pran}}\left(\boldsymbol{\nabla}\overline{\theta}_l\right)^2 \label{eq:resolved_temp_diss},\\
\boldsymbol{J}_l^\theta&=\overline{\boldsymbol u}_le^\theta_l-\frac{1}{\sqrt{\Ra\Pran}}\boldsymbol{\nabla} e^\theta_l+\boldsymbol \gamma_l\overline{\theta}_l \label{eq:resolved_temp_flux},\\
\text{and} \quad
\Pi_l^\theta&=-\boldsymbol{\gamma}_l\boldsymbol{\cdot\nabla}\overline{\theta}_l\label{eq:resolved_temp_transfer} . 
\end{align}
Equation \eqref{eq:resolved_temp_diss} describes the direct thermal dissipation of the resolved scales, \eqref{eq:resolved_temp_flux} the spatial redistribution of temperature variance and \eqref{eq:resolved_temp_transfer} the transfer rate between resolved and unresolved scales. We will refer to the latter as the heat transfer in the following.

\subsubsection{Averaged resolved temperature variance budget}

As before, we consider the time- and volume-averaged budget
\begin{align}
\avg{\chi_l}+\avg{\Pi_l^\theta}=\frac{1}{\sqrt{\Ra\Pran}}\Nu=\avg{\chi}, \label{eq:tempvar_l_global}
\end{align}
see Appendix \ref{app:tempbudget} for the derivation.
This budget shows that the total heat transport is balanced by the direct thermal dissipation and the heat transfer between scales. Because $\avg{\chi_l}\leq\avg{\chi}$, the averaged heat transfer between scales is down-scale, i.e.~$\avg{\Pi_l^\theta}>0$. This is consistent with classical theories, in which a direct temperature variance cascade is proposed \citep{lohse10arofm}.
The horizontally averaged budget is given by
\begin{align}
\avgA{\boldsymbol{\nabla\cdot J}_l^\theta}=-\avgA{\chi_l}-\avgA{\Pi_l^\theta} \label{eq:tempvar_l_horiz},
\end{align}
which shows that the spatial redistribution of the resolved temperature variance is balanced by the direct thermal dissipation and the heat transfer between scales.

\subsection{Numerical simulations}
{\renewcommand{\arraystretch}{1.2}
	\newcolumntype{G}{>{\columncolor{gray!30}}c}
	\newcolumntype{g}{>{\columncolor{gray!10}}c}
	\newcolumntype{C}{>{\columncolor{ProcessBlue!50}}c}

	\begin{table}
		\centering
		\begin{tabular}{g g G G G G G G g g g} 
			\multicolumn{2}{g}{Input} & 	\multicolumn{6}{G}{Output} &\multicolumn{3}{g}{Time scales}\\ \midrule
			$\Ra$  & $N_xN_y N_z$ & $\Nu$ & $\Nu_\varepsilon$ & $\Nu_\chi$ & $\Rey$ & $\lambda_s$ &$l_s$ & $T_t$ &$\tau$&$t_s$ \\ \midrule 
			$1.03\times10^4$ & $448^2\times 64$ &2.26 & 2.26 & 2.26 & 17.8 & 4.8& 2.4 & 1954 &1303&91\\
			$5.01\times10^4$& $768^2\times96$ &3.55 &3.55& 3.55 & 47.3 & 4.8& 2.4& 1092& 728&76\\
			$1.02\times10^5$& $1280^2\times140$ & 4.36 & 4.37 & 4.36& 69.2 & 4.8&2.4 & 701&467&74\\ 
			$1.03\times10^6$ &$2560^2\times208$ & 8.37 & 8.38 & 8.37  & 222.6& 4.8&2.4&752&451&73\\ 
			$1.07\times10^7$ & $3200^2\times256$&  16.04 & 16.06& 16.04 & 685.9 & 6.0&3.0 & 1151 & 765& 90\\ 
			$1.04\times10^8$ & $7200^2\times416$&  30.95 & 30.99 & 30.94 & 2004.2 & 6.0&3.0 & 362 & 196 & 96\\ 
		\end{tabular}
		\caption{Input and reference output parameters of the simulations with $\Pran=1$. The number of grid points in the vertical direction is $N_z$ and in the horizontal directions $N_x$ and $N_y$. $\Nu,~\Nu_\chi$ and $\Nu_\varepsilon$ are Nusselt numbers calculated based on the thermal driving, thermal and viscous dissipation, respectively. Here, the Reynolds number $\Rey=\sqrt{\avg{ \boldsymbol{u}^2} \Ra/\Pran}$ is based on the root-mean-square velocity. Additionally, $\lambda_s$ characterizes the wavelength of the turbulent superstructures, which is determined from the cross-spectrum of $u_z$ and $\theta$, and $l_s$ represents the filter width to separate the superstructures from turbulent fluctuations. Lastly, $T_t$ is the total runtime, $\tau$ the time window over which the averages are taken after the initial transient, and $t_s$ the characteristic time scale of the evolution of the superstructures. We adopt the definition of $t_s$ from \citet{pandey18natcomm} but base it on $\lambda_s$.}\label{tab:setup}
	\end{table}

The OBEs \eqref{eq:OBE} are solved numerically, using a compact sixth-order finite-difference scheme in space and a fourth-order Runge-Kutta scheme for time stepping \citep{lomax01}. The grid is non-uniform in the vertical direction for $\Ra\geq5\times10^4$, with monotonically decreasing grid spacing towards the wall. The pressure equation is solved with a factorization of the Fourier-transformed Poisson equation to satisfy the solenoidal constraint \citep{mellado12zamm}. The filter used in our analysis is implemented using a trapezoidal rule. The code is also freely available at \textit{https://github.com/turbulencia/tlab}.

We study the Rayleigh number regime from $\Ra\approx10^4$ up to $\Ra\approx10^8$ in a large aspect ratio domain with $\Gamma\approx24$ for $\Pran=1$. The full simulation details are provided in table \ref{tab:setup}. The Nusselt numbers shown are calculated based on the thermal driving $\Nu=\sqrt{\Ra\Pran}\avg{ u_z\theta }+1$, the viscous dissipation $\Nu_\varepsilon=\sqrt{\Ra\Pran}\avg{ \varepsilon}+1$ and the thermal dissipation $\Nu_\chi=\sqrt{\Ra\Pran}\avg{ \chi}$. Their mutual consistency serves as a resolution check of the simulations \citep{verzicco03jfm}. For our simulations, the different Nusselt numbers agree to $\unit[99]{\%}$ or better. Furthermore, the resolution requirements have been estimated \textit{a priori} as proposed in \citet{shishkina10njp}, and the relevant scale, i.e.~the Kolmogorov scale $\eta$ for $\Pran=1$, has been compared to the grid resolution \textit{a posteriori}.
In all cases we find that the maximum grid step $h$ is smaller than the Kolmogorov scale $\eta$, and that the vertical grid spacing $\Delta z$ is smaller than the height-dependent Kolmogorov scale based on $\avgA{\varepsilon}$ at the corresponding height. Together with the consistency of the Nusselt number, this shows that our simulations are sufficiently resolved. Further resolution studies can be found in \citet{mellado12jfm}. As a test for stationarity, we computed all terms in \eqref{eq:resolved_budget_global} and \eqref{eq:tempvar_l_global} individually. We find from our simulations that the left-hand sides agree with the right-hand sides to $\unit[99]{\%}$ for all considered filter widths.

\section{Results}
\label{sec:results}
 
In the following, we present numerical results to examine the scale dependence of the resolved energy budget as well as the resolved temperature variance budget.
We focus on the scale of the superstructures, for which we first have to characterize their scale.

\subsection{Determining the superstructure scale}
\label{sec:superstructure_scale}
In order to extract the length scale of the superstructures, we compute azimuthally and time-averaged spectra in horizontal planes (cf.~\citet{hartlep03prl, pandey18natcomm} and \citet{stevens18prf}). Specifically, we choose the azimuthally averaged cross-spectrum $E_{\theta u_z}(k)$ of the vertical velocity and the temperature in the midplane for the definition of the superstructure scale \citep{hartlep03prl}.
Here, $E_{\theta u_z}(k)$ is normalized in such a way that it integrates to $\avgA{Q}(z=0.5)$.
A representative example is shown in figure \ref{fig:crossspec}.
The peak of the spectrum characterizes the wavelength of the superstructures $\lambda_s=2\pi/k_{\lambda_s}$. The corresponding length scale $\lambda_s$ is listed in table \ref{tab:setup} for all simulations. The wavelength increases compared to the theoretical expectation for onset $\lambda_0=2.016$ \citep{getling98} and is largest for the highest Rayleigh numbers. The observed length scales are comparable with the ones obtained in previous studies of superstructures \citep{hartlep03prl,vonHardenberg08pra,stevens18prf,pandey18natcomm,fodor19blm}. 
Since a superstructure consists of a pair of a warm updraft and a cold downdraft, we choose the filter width $l_s\approx\lambda_s/2$ to investigate the energy and temperature variance budgets at the scale of the superstructure.
The values are given in table \ref{tab:setup}. We tested that small variations do not affect the outcome significantly. With this choice the individual large-scale up- and downdrafts are retained and the small-scale fluctuations are removed. We can then use \eqref{eq:resolved_budget} and \eqref{eq:tempvar_l} to characterize the energetics of the large-scale convection rolls and the associated superstructures and filter out the smaller-scale fluctuations.

Previous studies indicated that the length scales for the temperature and velocity field differ at high Rayleigh numbers \citep{pandey18natcomm,stevens18prf} when they are determined from the peak in the corresponding spectrum. However, recently \citet{krug19arxiv} studied linear coherence spectra of the vertical velocity and temperature field to argue that superstructures of the same size exist in both fields for $\Pran=1$ also at high $\Ra$. They found that the resulting scale essentially coincides with the peak of the cross-spectrum, which justifies the use of a single length scale for both fields. Note also that we use a single filter scale for all heights. This can be justified from the fact that the size of the superstructures does not noticeably vary with height and is closely connected to characteristic large scales close to the wall \citep{parodi04prl,vonHardenberg08pra,pandey18natcomm,stevens18prf,krug19arxiv}. The spectra of the temperature and the heat flux have a second maximum at larger wavenumbers close to the wall, which characterize smaller-scale fluctuations \citep{kaimal76jas,mellado16blm,krug19arxiv}. For completeness, we discuss the choice of the superstructure scale in more detail in Appendix \ref{app:spectra}.

\begin{figure}
		\centering
		\includegraphics[width=.5\textwidth]{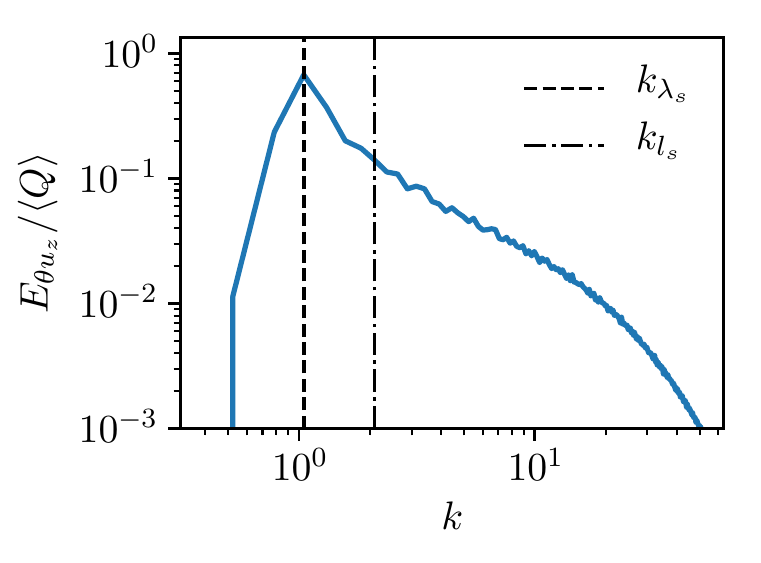} 
	\caption{Cross spectrum of the temperature and the vertical velocity in the midplane for $\Ra=1.07\times10^7$. The maximum wavenumber, which characterizes the large-scale rolls, is highlighted by the dashed line. The filter width (dash-dotted line) to separate superstructures and small-scale fluctuations is given by $k_{l_s}\approx2k_{\lambda_s}$. This choice removes the small-scale fluctuations and preserves the large-scale hot updrafts and cold downdrafts, which form the superstructures.
	The corresponding wavelengths are indicated in the snapshot of the temperature field in the midplane in figure \ref{fig:RBC_LSS}b.}
	\label{fig:crossspec}
\end{figure}

\subsection{Volume-averaged resolved energy budget}\label{sec:results_volume}
In this section, we study the volume-averaged resolved energy budget. We first consider a wide range of filter widths before focusing on the specific scale of the superstructures. We begin our discussion with the scale dependence of the stationary resolved energy budget \eqref{eq:resolved_budget_global}.
\begin{figure}
	\begin{subfigure}{0.5\linewidth}
		\centering
			\caption{}
		\includegraphics[width=1\textwidth]{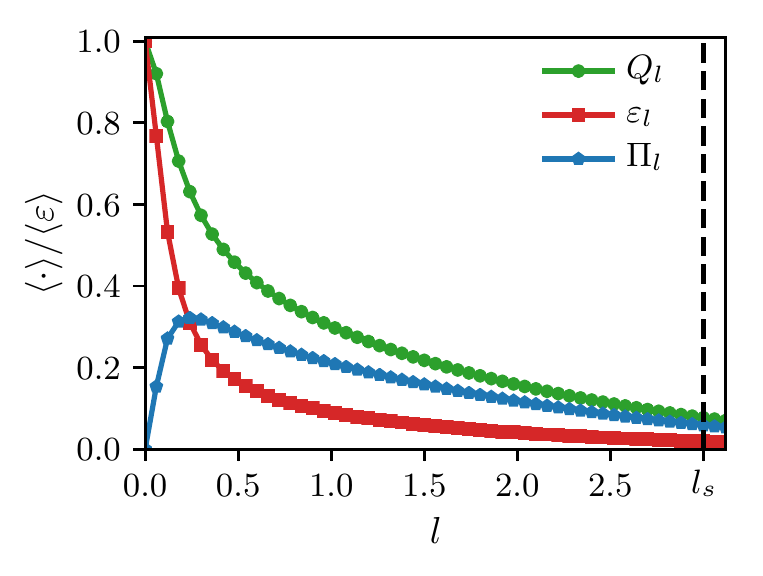} 
		\label{fig:energy_budget_global_a}
	\end{subfigure}
	\begin{subfigure}{0.5\linewidth}
		\centering
		\caption{}
		\includegraphics[width=1\textwidth]{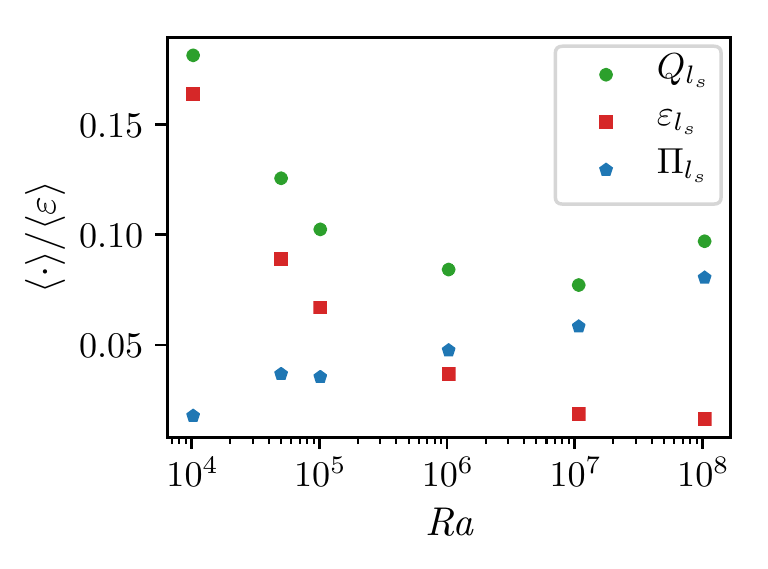} 
		\label{fig:energy_budget_global_c}
	\end{subfigure}
	\begin{subfigure}{0.5\linewidth}
		\centering
		\caption{}
		\includegraphics[width=1\textwidth]{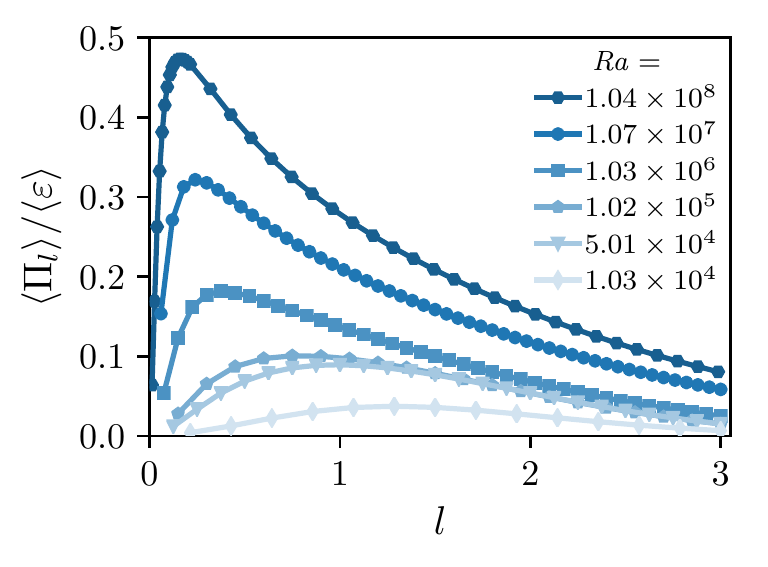} 
		\label{fig:energy_budget_global_b}
	\end{subfigure}
				\begin{subfigure}{0.5\linewidth}
					\centering
						\caption{}
					\includegraphics[width=1\textwidth]{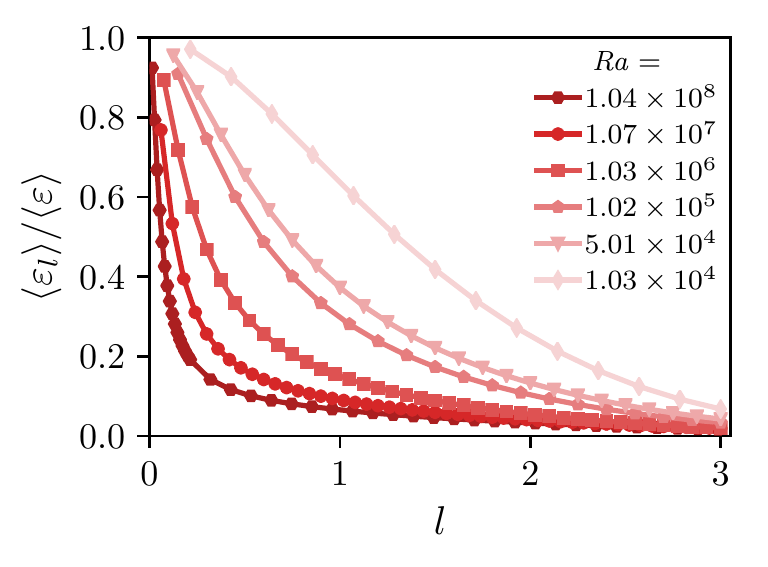} 
					\label{fig:energy_budget_global_b2}
				\end{subfigure}
					\begin{subfigure}{0.5\linewidth}
						\centering
								\caption{}
						\includegraphics[width=1\textwidth]{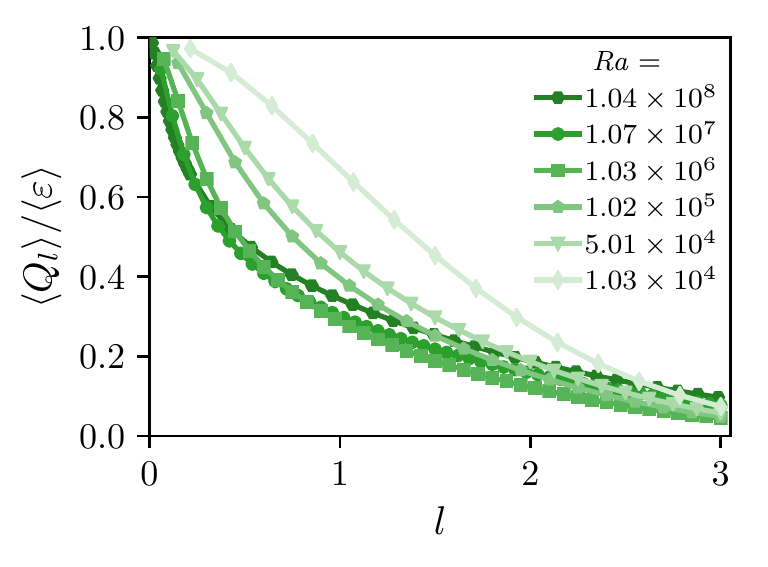} 
						\label{fig:energy_budget_global_b1}
					\end{subfigure}
					\begin{subfigure}{0.5\linewidth}
						\centering
							\caption{}
						\includegraphics[width=1\textwidth]{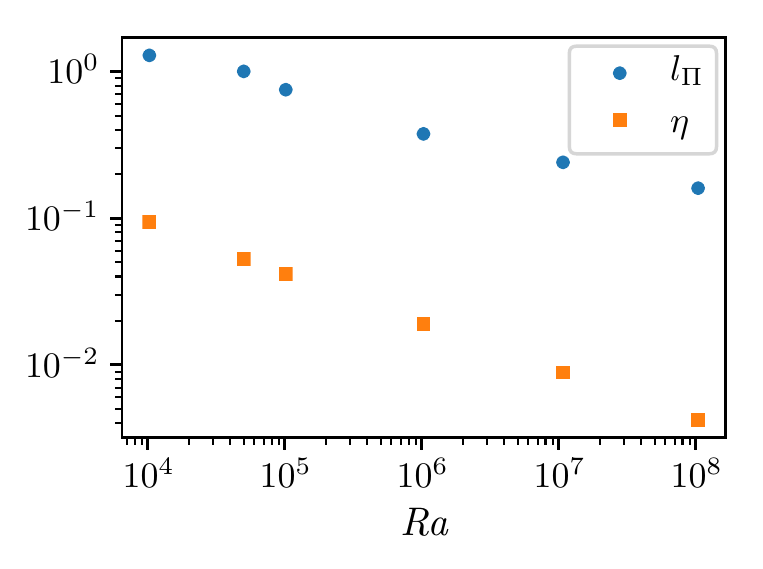} 
						\label{fig:energy_budget_global_d}
					\end{subfigure}
	\caption{(a) Contributions to the volume-averaged resolved energy budget for a range of filter scales $l$ for $\Ra=1.07\times10^7$. (b) Different contributions to the budget \eqref{eq:resolved_budget_global} at the superstructure scale $l_s$ as a function of $\Ra$. (c,d,e) Comparison of $\avg{ \Pi_l}$, $\avg{ \varepsilon_l}$ and $\avg{Q_l}$ for different $\Ra$. (f) Scale $l_\Pi$ of the maximum of $\avg{\Pi_l}$ compared to the Kolmogorov scale $\eta$ as a function of $\Ra$.}
	\label{fig:energy_budget_global}
\end{figure}
The different contributions are shown in figure \ref{fig:energy_budget_global_a} as a function of the filter width for $\Ra=1.07\times10^7$. The average energy input into the resolved scales $\avg{Q_l}$ and the direct dissipation $\avg{\varepsilon_l}$ decrease monotonically with increasing $l$. In contrast to that, the average energy transfer $\avg{\Pi_l}$ has a maximum at intermediate scales. For all shown filter widths $\avg{ \Pi_l } >0$, i.e.~the energy transfer acts on average as an energy sink as expected for three-dimensional turbulence (see discussion in section \ref{sec:mean_energy_budget}). In other words, there is a net energy transfer from the large to the small scales. 

How can we understand the functional form of $\avg{\Pi_l}$? At large scales, dissipation is comparably small and the energy transfer primarily balances the resolved thermal driving. With decreasing filter scale the energy input through thermal driving accumulates, which is why it increases with decreasing filter width. It is mostly balanced by the energy transfer, which increases accordingly. When the filter scale reaches the dissipative regime, the direct dissipation $\avg{\varepsilon_l}$ begins to dominate, and the energy transfer starts to decay and finally vanishes at $l=0$, as expected from the analytical limits derived above. The functional form of the energy transfer at small filter width is comparable to three-dimensional turbulence, see, e.g., \citet{ballouz18jfm,buzzicotti18jot}. Notably, at the superstructure scale $l_s$, only a small fraction, roughly $\unit[8]{\%}$, of the total energy input $\avg{Q}$ 
is injected into the resolved scales. Out of that approximately $\unit[76]{\%}$ are transferred to unresolved scales, and approximately $\unit[24]{\%}$ are directly dissipated. 

In figure \ref{fig:energy_budget_global_c}, we compare $\avg{\Pi_{l_s}}$, $\avg{Q_{l_s}}$ and $\avg{\varepsilon_{l_s}}$, respectively, at the scale of the superstructure $l_s$ for different Rayleigh numbers. 
The energy transfer becomes increasingly important compared to the direct dissipation at larger Rayleigh numbers. For $\Ra\geq1.07\times 10^7$ it is of the same order as the energy input, hence being crucially important for the energy budget of the turbulent superstructures.
We associate the relative increase of the energy transfer to an increase in turbulence for higher $\Ra$. 

Figures \ref{fig:energy_budget_global_b}, \ref{fig:energy_budget_global_b2} and \ref{fig:energy_budget_global_b1} show $\avg{\Pi_{l}}$, $\avg{Q_{l}}$ and $\avg{\varepsilon_{l}}$ as a function of filter width. In general, the energy transfer between scales acts as a sink and increases with $\Ra$, see figure \ref{fig:energy_budget_global_b}. In contrast, the direct dissipation decreases, see figure \ref{fig:energy_budget_global_b2}, for all considered scales. For the resolved energy input we do not observe simple trends, see figure \ref{fig:energy_budget_global_b1}. It is more constrained to small scales, yet there is still a non-vanishing energy input into the largest scales.

The scale $l_\Pi$ at which $\avg{ \Pi_l }$ is maximal decreases with $\Ra$, as shown in figure \ref{fig:energy_budget_global_d}. We expect this to be related to the shift of the dissipative range to smaller scales with increasing $\Ra$, since the energy transfer decays when the filter scale reaches the dissipative regime. The Kolmogorov scale $\eta$ characterizes the dissipative scale. As shown in figure \ref{fig:energy_budget_global_d}, $l_\Pi$ follows a similar trend as $\eta$.

\subsection{Horizontally averaged resolved energy budget}\label{sec:results_profiles}
In RBC the flow in the boundary layers and the bulk region is qualitatively different, as long as the boundary layers are not fully turbulent \citep{ahlers09rmp,lohse10arofm,chilla12epje}. To analyse the difference between these distinct regions in the resolved energy budget, we present results for the horizontally averaged energy budget \eqref{eq:resolved_budget_profile}.
This helps to understand the role of the boundary layers for the different contributions of the resolved energy budget in more detail.
Compared to the volume-averaged budget, there is an additional spatial flux term $\avgA{ \boldsymbol{\nabla \cdot J}_l }$, which redistributes energy vertically. The profiles of all the height-dependent contributions of \eqref{eq:resolved_budget_profile} at the superstructure scale $l_s$ are presented in figure \ref{fig:plane_budget}a for a simulation with $\Ra =1.07 \times 10^7$ as an example from the turbulent regime. They are compared to the unfiltered profiles in \ref{fig:plane_budget}b. The shown flux terms are calculated from the right-hand sides of \eqref{eq:resolved_budget_profile} and \eqref{eq:energy_budget}, respectively.
\begin{figure}
	\centering
\includegraphics[width=1\textwidth]{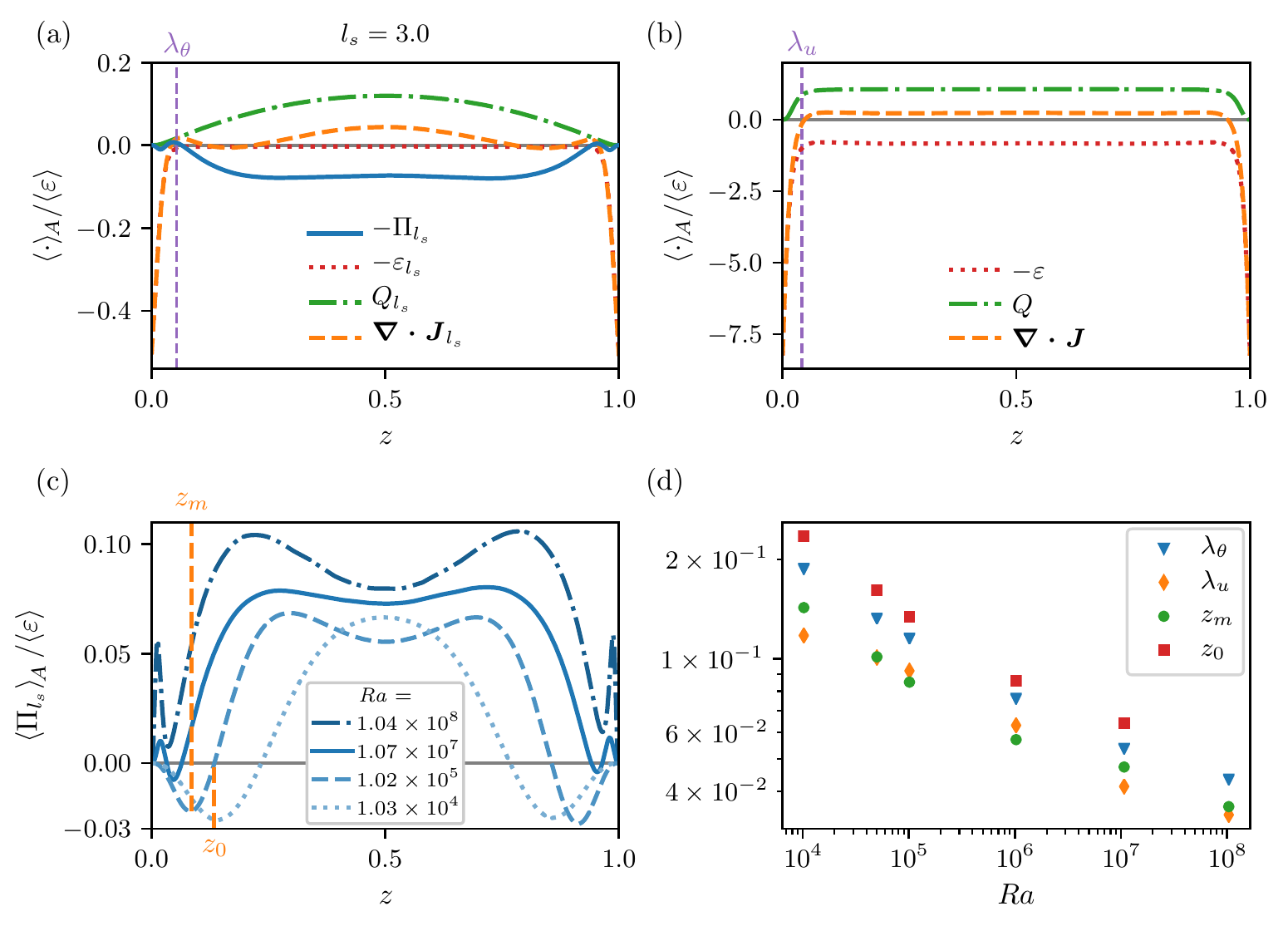} 
	\caption{(a) Different contributions to the horizontally averaged resolved energy budget at the superstructure scale $l_s$ and (b) unfiltered energy budget for $\Ra=1.07\times 10^7$ normalized by the total dissipation. (c) Energy transfer term at $l_s$ for different $\Ra$ normalized by the corresponding total dissipation. (d) Comparison of the distance from the wall to the first minimum $z_m$ and zero crossing $z_0$ of $\avgA{\Pi_{l_s}}$ with the boundary layer thicknesses of the temperature and the velocity fields as a function of $\Ra$.}
	\label{fig:plane_budget} 
\end{figure} 
The energy input into the resolved scales takes place mainly in the bulk and decays towards the wall. In contrast,
the direct dissipation primarily occurs near the wall and decays towards the bulk. 
The energy transfer is positive in a layer in the bulk, i.e.~it acts as a sink. Therefore, it effectively increases the dissipation, as it does for the volume-averaged balance. However, we also find an inverse energy transfer from the unresolved to the resolved scales near the wall in agreement with previous results for RBC \citep{togni15jfm,togni19jfm} and other wall-bounded flows \citep{domaradzki94pof,marati04jfm,cimarelli11jfm,cimarelli12pof,cimarelli15jfm,bauer19prf}.

A comparison of the energy transfer profiles for different Rayleigh numbers (see figure \ref{fig:plane_budget}c) shows that their form depends strongly on $\Ra$.
The energy transfer peaks always in the bulk and is exclusively a sink in this region, i.e.~it acts as an additional dissipation. Thus the bulk determines the behaviour of the volume-averaged energy transfer. With increasing $\Ra$ the width of the plateau of $\avgA{\Pi_{l_s}}$ in the bulk increases. 
For $\Ra<10^7$ the energy transfer close to the wall is characterized by a negative minimum, which means that there is a near-wall layer contributing to the driving of the resolved scales. 
With increasing $\Ra$ the near-wall structure of $\avgA{\Pi_{l_s}}$ changes and the inverse layer vanishes at the largest Rayleigh number. Here, it turns into a positive minimum. 
However, locally there are still regions of upscale transfer present. This illustrates that the boundary layers play a different role for the dynamics of the superstructures than the bulk.
We present an interpretation of this layer structure in terms of the plume dynamics in section \ref{sec:TransferAndPlumes}.
Note that the profiles are scale dependent, particularly at high $\Ra$. Therefore, the energy transfer close to the wall depends on the considered filter scale as well as the Rayleigh number and has to be interpreted carefully for this reason. We present a description of the dependence on the filter scale $l$
in Appendix \ref{app:Pi_l_profile}.

We shall make the first attempt to link the scale-resolved layer structure revealed in figure \ref{fig:plane_budget}c with the boundary layer structure of RBC. Figure \ref{fig:plane_budget}d shows the thickness of the thermal dissipation layer $\lambda_\theta$ and viscous dissipation layer $\lambda_u$ as a function of $\Ra$. The layers are defined as the distance to the wall at which the horizontally averaged thermal, respectively viscous, dissipation equals its volume average. \citet{petschel13prl} originally introduced these layers to study and compare boundary layers for different boundary conditions in RBC. They also compared their scaling as a function of $\Pran$ with classical boundary layer definitions. 
For an investigation of the Rayleigh number dependence of the different boundary layers we refer to \citet{scheel14jfm}, who showed that the scaling of the dissipation-based boundary layers differ from the classical ones. The boundary layers are indicated in figure \ref{fig:plane_budget}a and \ref{fig:plane_budget}b to present their relative position compared to the profiles.
In figure \ref{fig:plane_budget}d, the distance of the first local minimum of $\avgA{\Pi_{l_s}}$ to the lower wall $z_m$ and that of the subsequent zero crossing to the lower wall $z_0$, where the transfer changes from inverse to direct, are presented as a function of $\Ra$. (They are also highlighted for clarity in figure \ref{fig:plane_budget}c for $\Ra=1.02\times10^5$.)
All scales decrease with increasing $\Ra$ and follow a similar trend. Interestingly, $z_m$
appears to be bounded by the thermal layer.
This means that the inverse energy transfer mostly happens inside the thermal boundary layer, i.e.~close to the wall. We associate the decrease of its extent with the well-known shrinking of the boundary layers \citep{ahlers09rmp,scheel14jfm}.
For the highest $\Ra$, the inverse transfer layer vanishes, which we will discuss in section \ref{sec:TransferAndPlumes}. The minimum at $z_m$ now describes a direct transfer in contrast to the smaller $\Ra$ but is still inside the thermal boundary layer. Overall, this shows that the differences in the flow between the bulk and close to the wall are also represented in the structure of the transfer term.

\subsection{Effective resolved dissipation and implications for reduced models} 
\begin{figure}
	\centering
	\includegraphics[width=.5\textwidth]{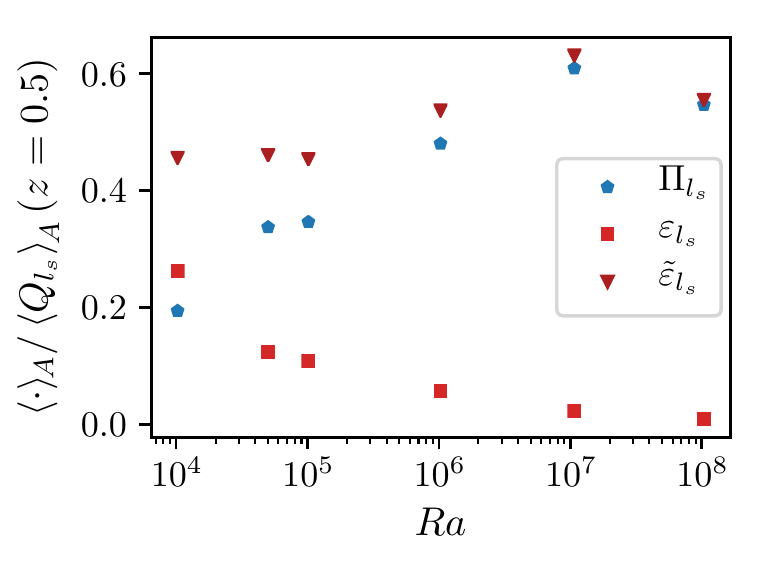} 
	\caption{Comparison of the direct dissipation $\varepsilon_{l_s}$, the energy transfer $\Pi_{l_s}$, and the effective dissipation $\tilde{\varepsilon}_{l_s}=\varepsilon_{l_s}+\Pi_{l_s}$ in the midplane normalized by the resolved energy input in the midplane as a function of $\Ra$. The energy transfer is significantly larger than the direct dissipation at high Rayleigh numbers.}
	\label{fig:plane_budget_renorma} 
\end{figure}
\citet{emran15jfm} and \citet{pandey18natcomm} have pointed out similarities between the turbulent superstructures and patterns close to the onset of convection. In this regime, analytical techniques are feasible \citep{bodenschatz00arofm}. Combined with the filtering approach, this could enable future developments of effective large-scale equations for RBC at high $\Ra$. To discuss these similarities and their implications, we draw comparisons between the resolved profiles at large $\Ra$ to the unfiltered profiles for a small $\Ra$ from the weakly nonlinear regime. As we have seen in the previous section, the energy transfer primarily contributes to the resolved energy budget as a sink term, resulting in an additional dissipation. We therefore consider the effective resolved dissipation $\tilde{\varepsilon}_l=\varepsilon_l+\Pi_l$ at the superstructure scale.
In figure \ref{fig:plane_budget_renorma}, the averaged effective dissipation in the midplane is shown as a function of $\Ra$ normalized by the resolved energy input in the midplane. We observe that the effective dissipation slightly increases until $\Ra=10^7$. It removes roughly half of the energy input in the midplane. The comparison with the energy transfer and the direct dissipation reveals that at high Rayleigh number, the transfer of energy to small scales is primarily responsible for the effective dissipation of the energy. The direct dissipation, in comparison, is negligible at high $\Ra$.

Figure \ref{fig:plane_budget_renorm} shows the resolved profiles at the scale of the superstructures compared to the unfiltered profiles from the weakly nonlinear regime.
\begin{figure}
		\includegraphics[width=1\textwidth]{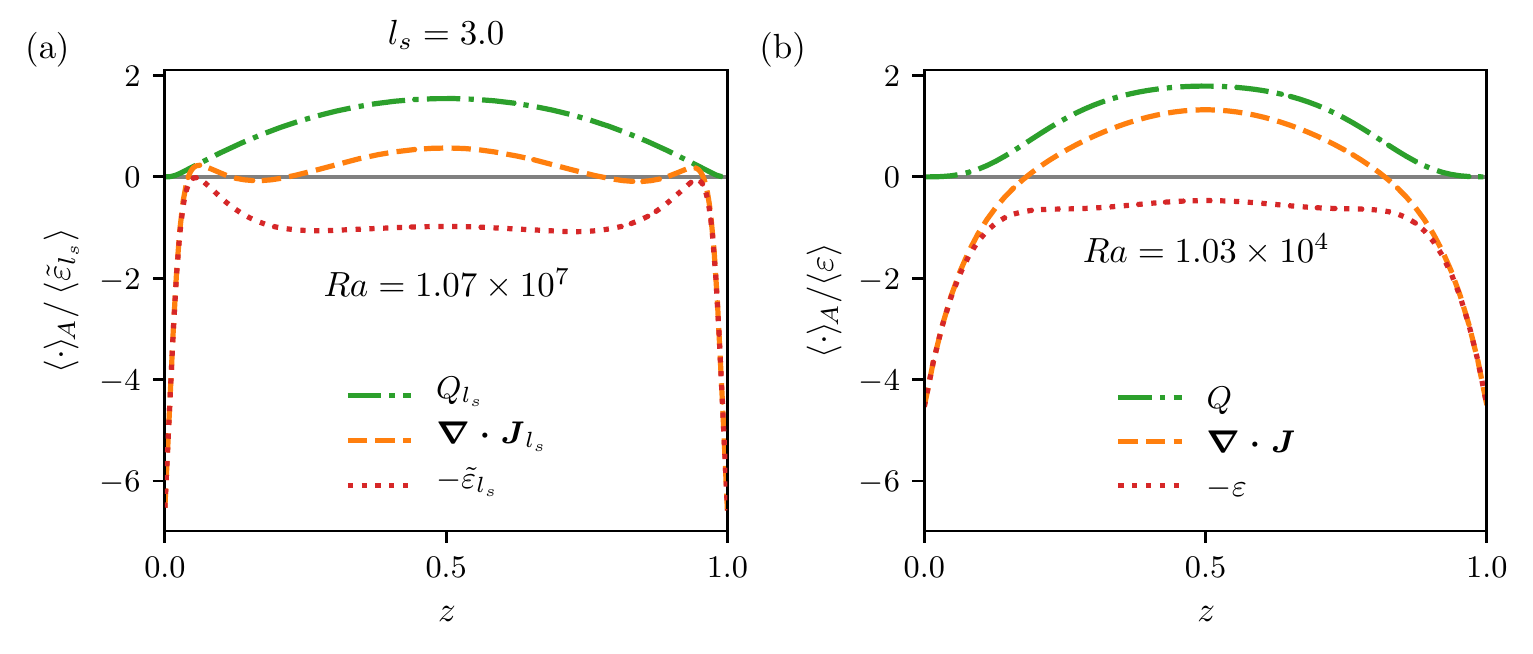} 
	\caption{(a) Resolved energy budget at the superstructure scale $l_s$ in terms of the effective dissipation $\tilde{\varepsilon}_{l_s}$ normalized with the total effective dissipation $\avg{ \tilde{\varepsilon}_{l_s}}$ for $\Ra=1.07\times 10^7$. (b) Unfiltered energy budget normalized with the total dissipation for $\Ra=1.03\times 10^4$, close to onset in the weakly nonlinear regime of convection.
	}
	\label{fig:plane_budget_renorm} 
\end{figure}
Close to the wall, the effective resolved dissipation and the redistribution differ from the corresponding profiles close to onset. Close to the midplane, the height-dependent profiles from the resolved budget and the original budget compare quite well, although some quantitative differences are visible. This indicates that an effective dissipation may capture the effect of the energy transfer on the superstructures in the bulk. The more complex near-wall behaviour of the superstructures at high $\Ra$ requires more elaborate approaches.
 
\subsection{Energy transfer rate and plume dynamics}
\label{sec:TransferAndPlumes}
In RBC plumes play a crucial role in the dynamics and are essential parts of the superstructures.
Using the filtering approach we can connect flow structures and their contribution to the energy budget. To gain insight into their role in the energy transfer, we discuss the local energy budget.
Figure \ref{fig:pi_l_theta} shows vertical cuts through the system for the energy transfer field and the temperature field for different $\Ra$. Especially in the weakly nonlinear regime, we observe a spatial correlation between plume impinging and detaching and the direction of the energy transfer. Regions of plume detachment correspond to regions of energy transfer to the unresolved scales, whereas regions of plume impinging correspond to regions of energy transfer from the small to the large scales. Similar observations have been made by \citet{togni15jfm}, who also found an inverse transfer from small to large scales connected to plume impinging. Due to the increasingly complex and three-dimensional motions at larger $\Ra$, see figure \ref{fig:pi_l_theta}b and \ref{fig:pi_l_theta}c, this spatial correlation is weakening. This is due to the fact that fewer plumes extend throughout the entire cell and are more likely to be deflected on their way from the top to the bottom plate or \textit{vice versa}. Hence they do not experience the sharp temperature gradient at the boundary layers. Instead, they release their temperature in the bulk and do not impinge on the boundary layers. This prevents the strong enlargement of individual plumes and the corresponding energy transfer to the large scales. However, clustered plumes, which effectively form large-scale plumes, still impinge on the walls and cause an inverse energy transfer. 
 \begin{figure}
 	\centering
   \includegraphics[width=1\textwidth]{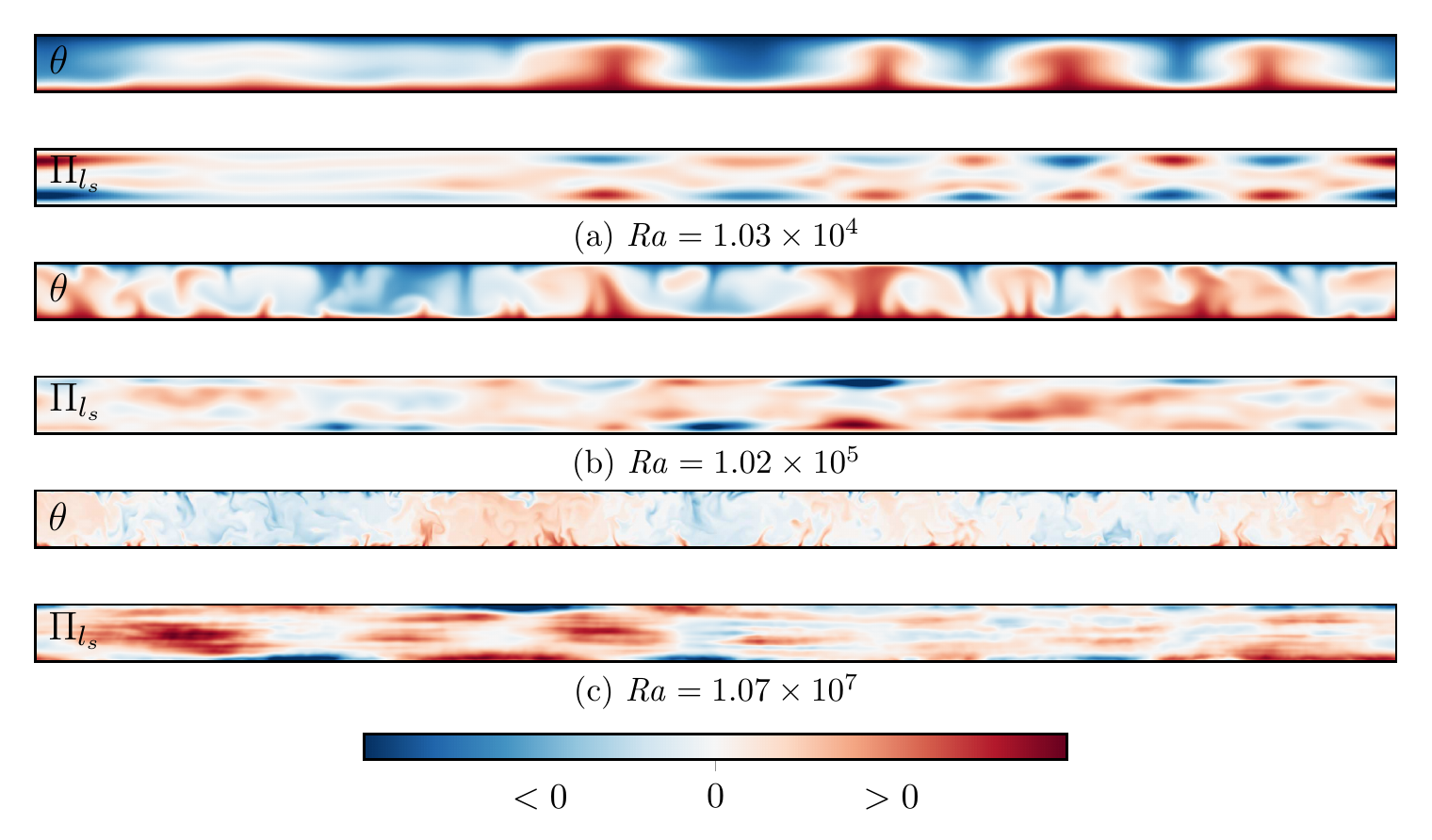} 
 \caption{ Comparison of an instantaneous snapshot of the temperature field and the energy transfer (normalized to unit maximum amplitude) between scales for (a) $\Ra=1.03\times10^4$, (b) $\Ra=1.02\times10^5$, and (c) $\Ra=1.07\times10^7$ at the superstructure scale $l_s$. Close to onset in the weakly nonlinear regime, a direct connection between the plume dynamics, i.e.~impinging and detachment, and the direction of the energy transfer is present. On impinging the plume heads enlarge, which is accompanied by an inverse energy transfer. During detachment the plumes shrink and there is a direct energy transfer.}
 \label{fig:pi_l_theta}
 \end{figure}
From the horizontally averaged energy transfer, see figure \ref{fig:plane_budget}c, we conclude that the inverse transfer caused by plume impinging exceeds the direct transfer caused by plume detaching, at least in the weakly nonlinear regime. However, at the largest Rayleigh number, the layer of inverse transfer vanishes. Here, the direct transfer caused by plume detaching exceeds the inverse transfer.

How can the above considerations be related to the findings for the volume-averaged energy budget? At small Rayleigh numbers the direct transfer in the bulk and the inverse transfer close to the wall almost balance, resulting on average in a small direct transfer. At larger $\Ra$ the inverse transfer caused by impinging is reduced because only a fraction of the released plumes reaches the opposite boundary layer. Here, the width of the inverse transfer layer is reduced. 
At the same time, the direct transfer increases and the corresponding layer becomes larger.
The direct transfer consequently grows on average with $\Ra$. For a discussion of the scale dependence of plume dynamics connected to the direction of the energy transfer see Appendix \ref{app:Pi_l_profile}. There, we discuss $\Pi_l$ and the profiles $\avgA{\Pi_l}$ for varying filter scales $l$.

 \subsection{Volume-averaged resolved temperature variance budget}\label{sec:results_volume_temp}
 \begin{figure}
 	\begin{subfigure}{.5\linewidth}
 		\centering
  		\caption{}
 		\includegraphics[width=1\textwidth]{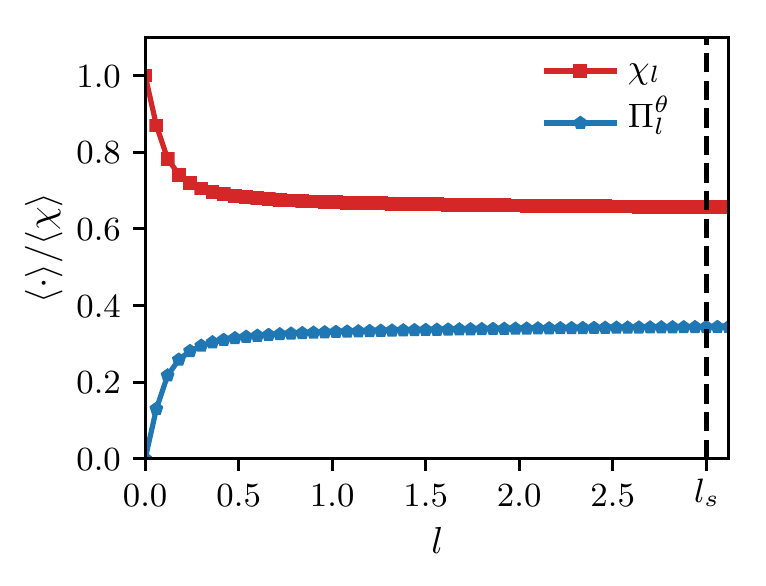}
 		\label{fig:pi_theta_l_a}
 	\end{subfigure}
  	\begin{subfigure}{.5\linewidth}
  		\centering
    \caption{}
  		\includegraphics[width=1\textwidth]{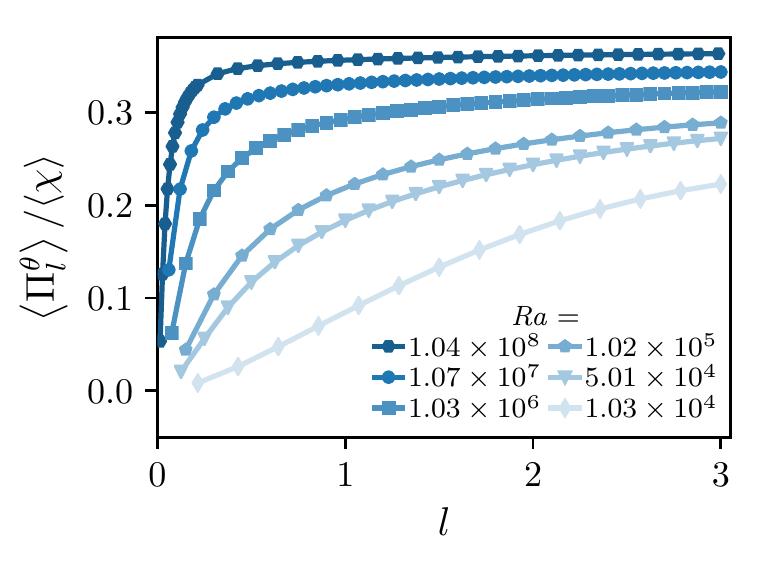}
  		\label{fig:pi_theta_l_b}
  	\end{subfigure}
 	\begin{subfigure}{.5\linewidth}
 		\centering
  		\caption{}
 		\includegraphics[width=1\textwidth]{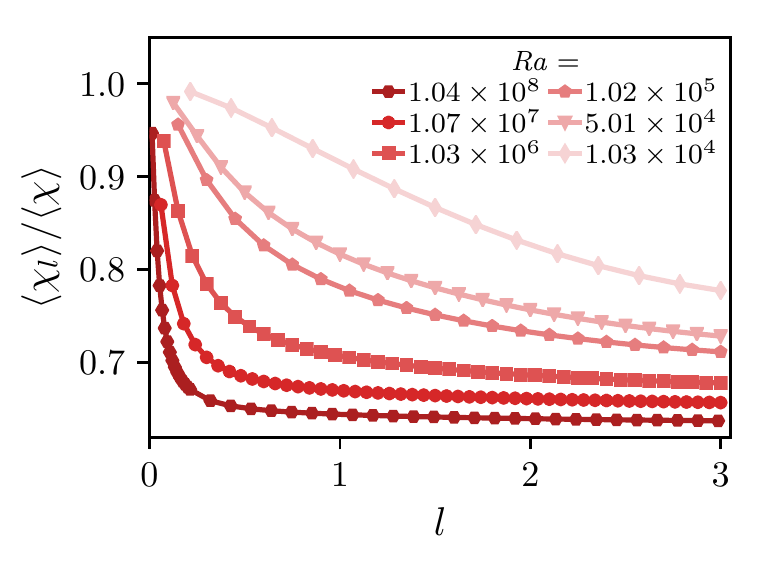} 
 		\label{fig:pi_theta_l_c}
 	\end{subfigure}
 	\begin{subfigure}{.5\linewidth}
 		\centering
  		\caption{}
 		\includegraphics[width=.975\textwidth]{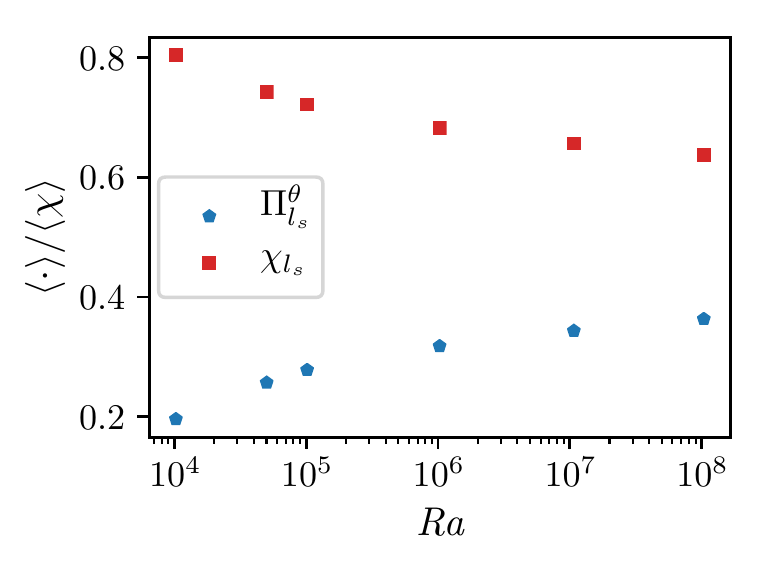} 
 		\label{fig:pi_theta_l_d}
 	\end{subfigure}
 	\caption{(a) Volume-averaged temperature variance budget for $\Ra=1.07\times10^7$. (b) Heat transfer $\avg{\Pi_l^\theta}$ and (c) thermal dissipation $\avg{\chi_l}$ for different $\Ra$ as a function of filter scale. (d) $\avg{\Pi^\theta_{l_s}}$ and $\avg{\chi_{l_s}}$ at the scale of the superstructures as a function of $\Ra$.}
 	\label{fig:pi_theta_l}
 \end{figure}
For completeness, we here complement the previous section with the consideration of the budget of the resolved temperature variance.
The balance \eqref{eq:tempvar_l_global} shows that the total thermal dissipation is split into two contributions: the resolved dissipation $\avg{\chi_l}$ and the heat transfer $\avg{\Pi_l^\theta}$. As illustrated in figure \ref{fig:pi_theta_l_a}, the resolved thermal dissipation exceeds the heat transfer at all scales, including the scale of the superstructure for the considered Rayleigh number. This is qualitatively different from the behaviour observed for the contributions to the kinetic energy balance. The heat transfer and direct dissipation both approach a constant value after an initial increase for small filter width. At these scales, they are approximately scale independent and the transfer of temperature variance is down-scale. This is important for the phenomenology of RBC.
In fact, both the Obukhov-Corrsin theory as well as the Bolgiano-Obukhov theory rest on a direct cascade picture for the temperature variance, consistent with our observations.
 A more detailed treatment of these considerations is beyond the scope of our work, and we refer the reader to \citet{lohse10arofm,ching14,verma17njp} and \citet{verma18} and references therein.
 Similarly to the energy transfer, the heat transfer increases with increasing $\Ra$ and the resolved thermal dissipation decreases, see figure \ref{fig:pi_theta_l_b}, \ref{fig:pi_theta_l_c}, and \ref{fig:pi_theta_l_d}. The heat transfer is always positive and, therefore, acts as a thermal dissipation for the resolved scales.
 
\subsection{Horizontally averaged resolved temperature variance budget}\label{sec:results_profiles_temp}
 \begin{figure}
 	 	 		\centering
 	 	 		\includegraphics[width=1\textwidth]{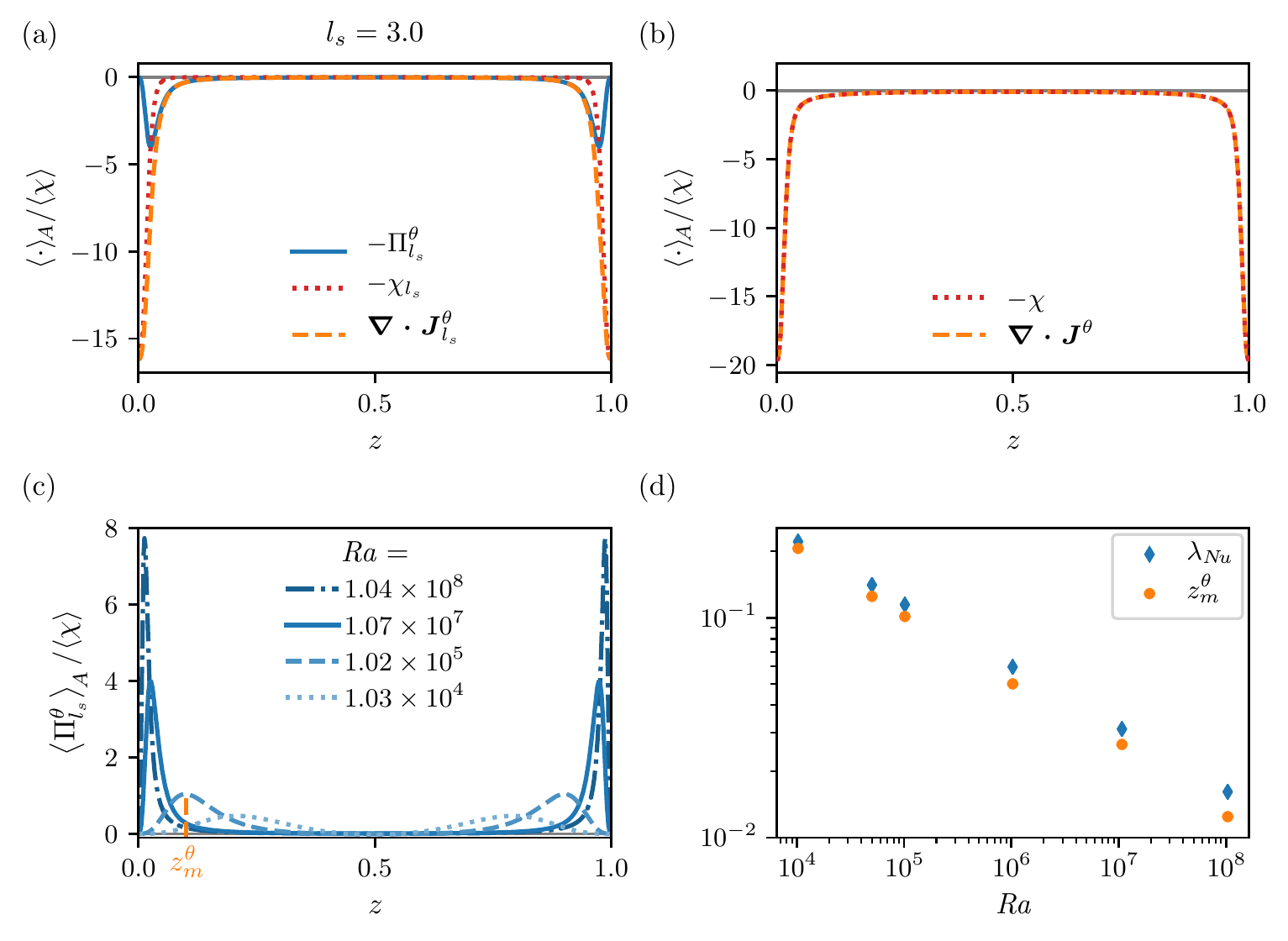} 
 	 	\caption{(a) Different contributions to the horizontally averaged resolved temperature variance budget at the superstructure scale $l_s$ and (b) unfiltered temperature variance budget for $\Ra=1.07\times 10^7$, normalized by the total thermal dissipation. (c) Profile of the heat transfer $\avgA{ \Pi^\theta_{l_s}}$ at the superstructure scale for different $\Ra$. (d) Distance $z_m^\theta$ from the wall to the maximum of $\avgA{ \Pi^\theta_{l_s}}$ compared to the thermal boundary layer thickness $\lambda_{N\!u}$ as function of $\Ra$. $z_m^\theta$ is also highlighted in (c) for $\Ra=1.02\times10^5$.}
 	 	\label{fig:thermal_variance_profiles}
 	 \end{figure}
The profiles of all the contributions to the horizontally averaged resolved temperature variance budget are shown in figure \ref{fig:thermal_variance_profiles}a for $\Ra=1.07\times10^7$ and compared to the unfiltered profiles in figure~\ref{fig:thermal_variance_profiles}b as an example from the turbulent regime. Here, the unfiltered flux is given by $\boldsymbol{J}^\theta=\boldsymbol {u}e^\theta-\boldsymbol{\nabla} e^\theta/\sqrt{\Ra\Pran}$, which can be obtained from \eqref{eq:resolved_temp_flux} in the limit of a vanishing filter width $l$. The resolved thermal dissipation follows a very similar form as the original thermal dissipation. It almost vanishes in the bulk and strongly increases towards the walls in the boundary layers. The heat transfer is positive for almost all heights and also vanishes in the bulk. It has a strong peak close to the walls and acts exclusively as a thermal dissipation. This is similar for different $\Ra$ as shown in figure \ref{fig:thermal_variance_profiles}c. A notable exception is at small $\Ra$, where it is slightly negative, i.e.~up-scale, close to the midplane. The peak of $\avgA{\Pi_{l_s}^\theta}$ increases in magnitude with increasing $\Ra$ and its distance to the wall $z_m^\theta$ decreases. The peak almost coincides with the height of the thermal boundary layer $\lambda_{N\!u}$, see figure \ref{fig:thermal_variance_profiles}d. 
In this region, the temperature variance deposited by the resolved heat flux is partly transferred to smaller scales and mainly dissipated.

Comparing the resolved energy with the resolved temperature variance budget, there are qualitatively similar scale dependencies. The transfers between scales increase with increasing $\Ra$ and act on average as a dissipation. However, the volume-averaged heat transfer is roughly constant after an initial increase at small scales, whereas the energy transfer decays after a maximum at small scales. At the scale of the superstructures, the volume-averaged heat transfer is smaller than the corresponding direct thermal dissipation for all $\Ra$. In contrast, the volume-averaged energy transfer exceeds the direct dissipation at large $\Ra$. Additionally, the profiles at the superstructure scale show qualitative differences, i.e.~the heat transfer is almost exclusively down-scale for all heights while the energy transfer shows a layer of up-scale energy transfer as well.

\section{Summary}
\label{sec:conclusions}

We investigated the scale-resolved kinetic energy and temperature variance budgets of RBC at Rayleigh numbers in the range $1.03\times10^4\leq\Ra\leq1.04\times 10^8$ for a fixed $\Pran=1$ and a high aspect ratio ($\Gamma\approx24$) with a focus on the interplay of turbulent superstructures and turbulent fluctuations. 
As a starting point, we generalized the volume-averaged kinetic energy and temperature variance budgets to scale-dependent budgets of the resolved fields. 
For the kinetic energy budget, this results in a balance between the resolved energy input, the direct large-scale dissipation and an energy transfer to the unresolved scales. It shows that the small-scale fluctuations play an important role for the energy balance of the large scales. For our simulations at the highest Rayleigh numbers under consideration, we find that the energy transfer to the smaller scales is of comparable magnitude to the resolved energy input at the superstructure scale. This means that the generation of small-scale turbulence acts as a dissipation channel for the large scales, which qualitatively confirms the classic picture that small-scale turbulence introduces an effective dissipation.

When resolving the energy transfer with respect to height, a more complex picture emerges which, in particular, reveals the role of the boundary layers. The height-dependent balance of the distinct terms is summarized in figure \ref{fig:energy_budget_sketch} at the superstructure scale. Panel (a) shows that most of the energy input due to thermal driving takes place in the bulk. From there, energy is transferred to smaller scales, see panel (b), and transported towards the wall, see panel (c).
While the direct large-scale dissipation is comparably small in the bulk, its main contribution stems from regions close to the wall, see panel (d). There, the situation is more complex. We find an additional inverse energy transfer from the small to the large scales for $\Ra<10^8$ and a minimum for the largest considered Rayleigh number. This illustrates that the boundary layers play a distinct role for the energy budget of the superstructures. 

\begin{figure}
	\centering
	\includegraphics[width=1\textwidth]{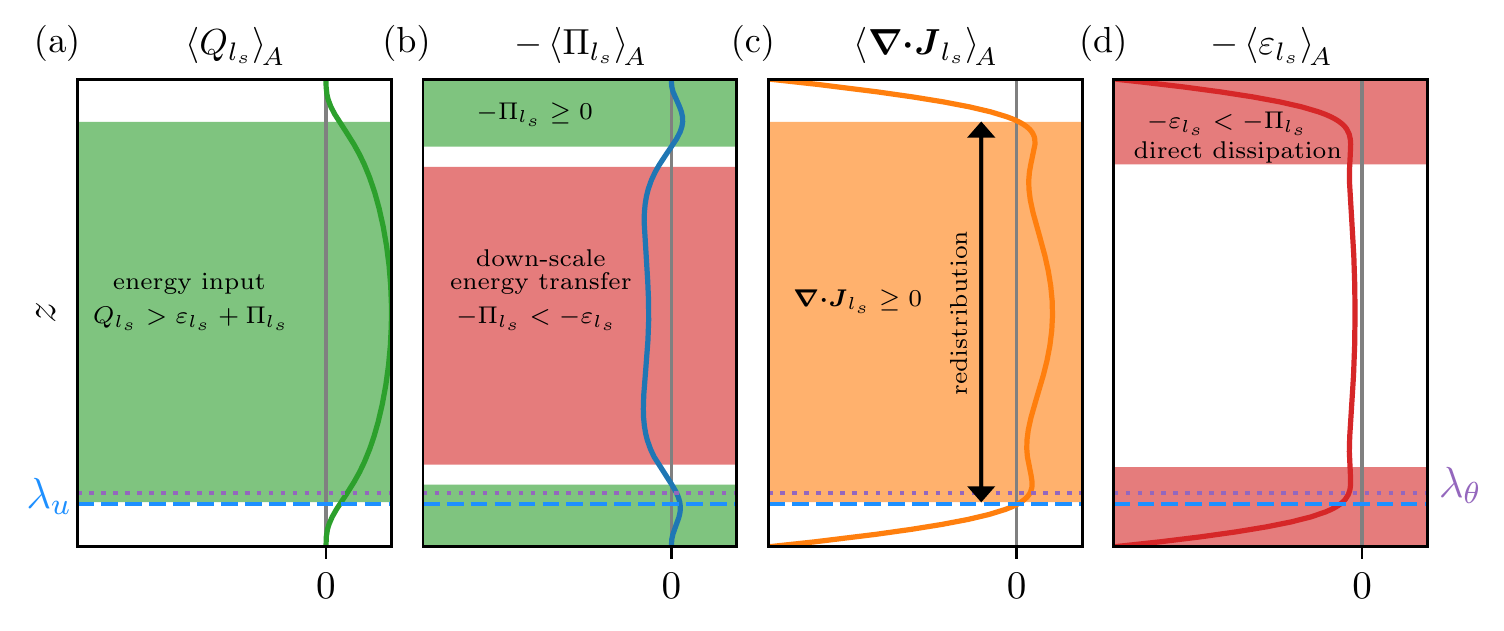}
	\caption{Sketch of the resolved energy balance at the scale of the superstructure, highlighting the distinct structure of the bulk and boundary layer. The profiles are obtained from a simulation at $\Ra=1.02\times10^5$ as an illustrative example for the moderately turbulent regime. The dissipation layer $\lambda_u$ and the thermal dissipation layer $\lambda_\theta$ are indicated by the dashed and dotted lines, respectively. Energy input regions are highlighted in green, direct dissipation and down-scale energy transfer in red, and spatial redistribution in orange.}
	\label{fig:energy_budget_sketch}
\end{figure}
Consistent with previous studies \citep{emran15jfm,pandey18natcomm}, we find qualitative similarities between the energy budget of turbulent superstructures and that of patterns in the weakly nonlinear regime. The resolved energy budget of the superstructures and the standard energy budget at the onset of convection show qualitative similarities in the midplane when the energy transfer to smaller scales is interpreted as an effective dissipation. This may open possibilities for modelling the large-scale structure of turbulent convection at high Rayleigh numbers. 

In order to gain insight into the origin of the inverse energy transfer, we studied the spatially resolved energy transfer. At small $\Ra$, there is a direct correspondence between plume impinging and plume detaching and the direction of the energy transfer. The enlargement of the plume head during impinging is accompanied by an energy transfer to the large scales. Conversely, the small scales are fed during plume detachment. A stronger inverse transfer caused by plume impinging can therefore result in the layer of inverse transfer observed close to the wall. However, in the turbulent regime, the lateral motion of the plumes is increased, which prevents the impinging on the boundary layers and the corresponding inverse energy transfer. Finally, at the largest Rayleigh number, the inverse layer vanishes.

We complemented the investigations of the resolved energy budget with the study of the resolved temperature variance budget. We find that the heat transfer between scales is roughly scale independent at large scales in the turbulent regime. Here, at the scale of the superstructures, the averaged direct thermal dissipation exceeds the averaged heat transfer for all considered $\Ra$. This is different from the behaviour of the energy transfer, and the direct thermal dissipation is more relevant for the balance of the temperature variance of the superstructures. Furthermore, the study of the height-dependent profiles showed that the heat transfer acts as a thermal dissipation at all heights for large Rayleigh numbers and is strongly peaked close to the boundary layers.

In summary, our investigations reveal the impact of turbulent fluctuations on the large-scale convection rolls in turbulent Rayleigh-B\'enard convection. In future investigations, it will be interesting to see whether the turbulent effects reach an asymptotic state at sufficiently high Reynolds numbers. This could open the possibility for universal effective large-scale models for Rayleigh-B\'enard convection at high Rayleigh numbers.

\begin{acknowledgments}
	\section*{Acknowledgments}
	This work is supported by the Priority Programme SPP 1881 Turbulent Superstructures of the Deutsche Forschungsgemeinschaft. 
	D.V.~gratefully acknowledges partial support by ERC grant No 787361-COBOM.
	Computational resources of the Max Planck Computing and Data Facility and support by the Max Planck Society are gratefully acknowledged.
\end{acknowledgments}
\section*{Declaration of Interests}
The authors report no conflict of interest.
\appendix
\captionsetup[subfigure]{singlelinecheck=on}
\section{Connection between volume-averaged resolved energy budget and original budget}\label{app:Gammal}
Under the assumptions that the filtered fields obey the same boundary conditions as the unfiltered ones, and that the filter preserves volume averages, the statistically stationary energy and temperature variance budgets can be related to the Nusselt number.
First, we can reformulate the resolved energy input 
\begin{align}
	\avg{\overline{u_z}_l\overline{\theta}_l}=\avg{\overline{\left(u_z\theta\right)}_l}-\avg{\boldsymbol{\gamma}_l\cdot\hat{\boldsymbol{z}}}=\avg{u_z\theta}-\avg{\boldsymbol{\gamma}_l\cdot\hat{\boldsymbol{z}}},
\end{align}
where we have used \eqref{eq:heat_flux} and that the filter preserves the volume average.
Then we find with $\Nu=\sqrt{\Ra\Pran}\avg{ u_z\theta}+1$ that
\begin{align*}
\avg{\overline{u_z}_l\overline{\theta}_l}=\avg{Q_l}=	\frac{1}{\sqrt{\Ra\Pran}}(\Nu-1)-\avg{\boldsymbol{\gamma}_l\cdot\hat{\boldsymbol{z}}} .
\end{align*}
This is inserted into \eqref{eq:resolved_budget_global}, and combined with \eqref{eq:EkinV} we obtain
\begin{align}
\avg{\varepsilon_l}+ \avg{\Pi_l}+\avg{ \boldsymbol{\gamma}_l\cdot\hat{\boldsymbol{z}}}&=\frac{1}{\sqrt{\Ra\Pran}}\left(\Nu-1\right)=\avg{\varepsilon} \label{eq:dissipation_l}.
\end{align}
This shows that the total kinetic energy dissipation $\avg{\varepsilon}$ is split into energy transfer between scales $\avg{\Pi_l}$,
direct dissipation of the resolved scales $\avg{\varepsilon_l}$ and the thermal driving of the unresolved scales $\avg{\boldsymbol{\gamma}_l\cdot\hat{\boldsymbol{z}}}$.
With \eqref{eq:resolved_budget_global} and \eqref{eq:EkinV}, we can write equation \eqref{eq:dissipation_l} also as
\begin{align}
\avg{ Q_l}+\avg{ \boldsymbol{\gamma}_l\cdot\hat{\boldsymbol{z}}}&=\frac{1}{\sqrt{\Ra\Pran}}\left(\Nu-1\right)=\avg{ Q},
\end{align}
in which the total energy input is split into the resolved energy input and the turbulent heat flux.
If we introduce the resolved Nusselt number $\Nu_l=\sqrt{\Ra\Pran}\avg{ Q_l }+1$, this relation can be written as
\begin{align}
\Nu_l+\sqrt{\Ra\Pran}\avg{\boldsymbol{\gamma}_l\cdot\hat{\boldsymbol{z}}}&=\Nu,
\end{align}
which shows that $\Nu$ is split into $\Nu_l$ and the heat flux into the unresolved scales $\sqrt{\Ra\Pran}\avg{ \boldsymbol{\gamma}_l\cdot\hat{\boldsymbol{z}}}$.

\section{Volume-averaged resolved temperature variance budget}\label{app:tempbudget}
Here we derive the volume-averaged resolved temperature variance budget \eqref{eq:tempvar_l_global}. We take the volume average of \eqref{eq:tempvar_l},
\begin{align}
\avg{\boldsymbol{\nabla\cdot J}_l^\theta}&=-\avg{\chi_l}-\avg{\Pi_l^\theta} ,
 \label{eq: b1}
\end{align}
in which the temporal derivative vanishes in the statistically stationary state. In contrast to the kinetic energy budget, the flux term does not vanish for the temperature variance. We obtain
\begin{align}
    \avg{\boldsymbol{\nabla\cdot J}_l^\theta}&=\avg{\boldsymbol{\nabla \cdot}\left(\overline{\boldsymbol{u}}_le^\theta_l-\frac{1}{\sqrt{\Ra\Pran}}\boldsymbol{\nabla} e^\theta_l+\boldsymbol{\gamma}_l\overline{\theta}_l\right)}=-\frac{1}{\sqrt{\Ra\Pran}}\avg{\nabla^2e^\theta_l},
    \label{eq: b2}
\end{align}
since the contributions containing $\overline{\boldsymbol{u}}_l$ vanish because of the boundary conditions.
To relate the flux term to the Nusselt number, we write the volume integral in the form
\begin{align}
\int_V \nabla^2e^\theta_l\,\mathrm{d}V=\int_V\boldsymbol{\nabla\cdot}\left(\overline{\theta}_l\boldsymbol{\nabla}\overline{\theta}_l\right)\,\mathrm{d}V=\int_{\partial V}\left(\overline{\theta}_l\boldsymbol{\nabla}\overline{\theta}_l\right)\cdot\hat{\boldsymbol{n}}\,\mathrm{d}A .
\end{align}
In the last integral the contributions from the sidewalls vanish because of the periodic boundary conditions. Therefore only the integration over the top and bottom wall remains, at which the temperature is constant, i.e.~$\overline{\theta}_l(z=0,1)=\pm1/2$. This gives
\begin{align}
  \int_{\partial V}\left(\overline{\theta}_l\boldsymbol{\nabla}\overline{\theta}_l\right)\cdot\hat{\boldsymbol{n}}\,\mathrm{d}A&=-\frac{1}{2}\left(\partial_z\int_{z=0} \theta\,\mathrm{d}A+\partial_z\int_{z=1}  \theta\,\mathrm{d}A\right),
\end{align}
where we used the fact that $\theta$ is constant at the top and bottom wall, and therefore $\overline{\theta}_l(z=0,1)=\theta(z=0,1)$.
The Nusselt number is defined as 
\begin{align}
  \Nu=\sqrt{\Ra\Pran}\avgA{u_z\theta}-\partial_z\avgA{\theta},
\end{align}
which is independent of $z$ (see, e.g.~\citet{scheel14jfm}). At the top and bottom wall $\boldsymbol{u}=0$ and
$\Nu(z=0,1)=-\partial_z\avgA{\theta}(z=0,1)$, and we find
\begin{align}
\avg{  \nabla^2e^\theta_l}=-\frac{1}{2}\left[\partial_z\avgA{\theta}(z=0)+\partial_z\avgA{\theta}(z=1)\right]=\Nu.
\end{align}
Substituting this back into \eqref{eq: b1} results in the volume-averaged balance \eqref{eq:tempvar_l_global}.
\section{Height-dependent spectra}\label{app:spectra}
\begin{figure}
	\centering
	\begin{subfigure}{0.475\linewidth}
		\centering
		\includegraphics[width=1\linewidth]{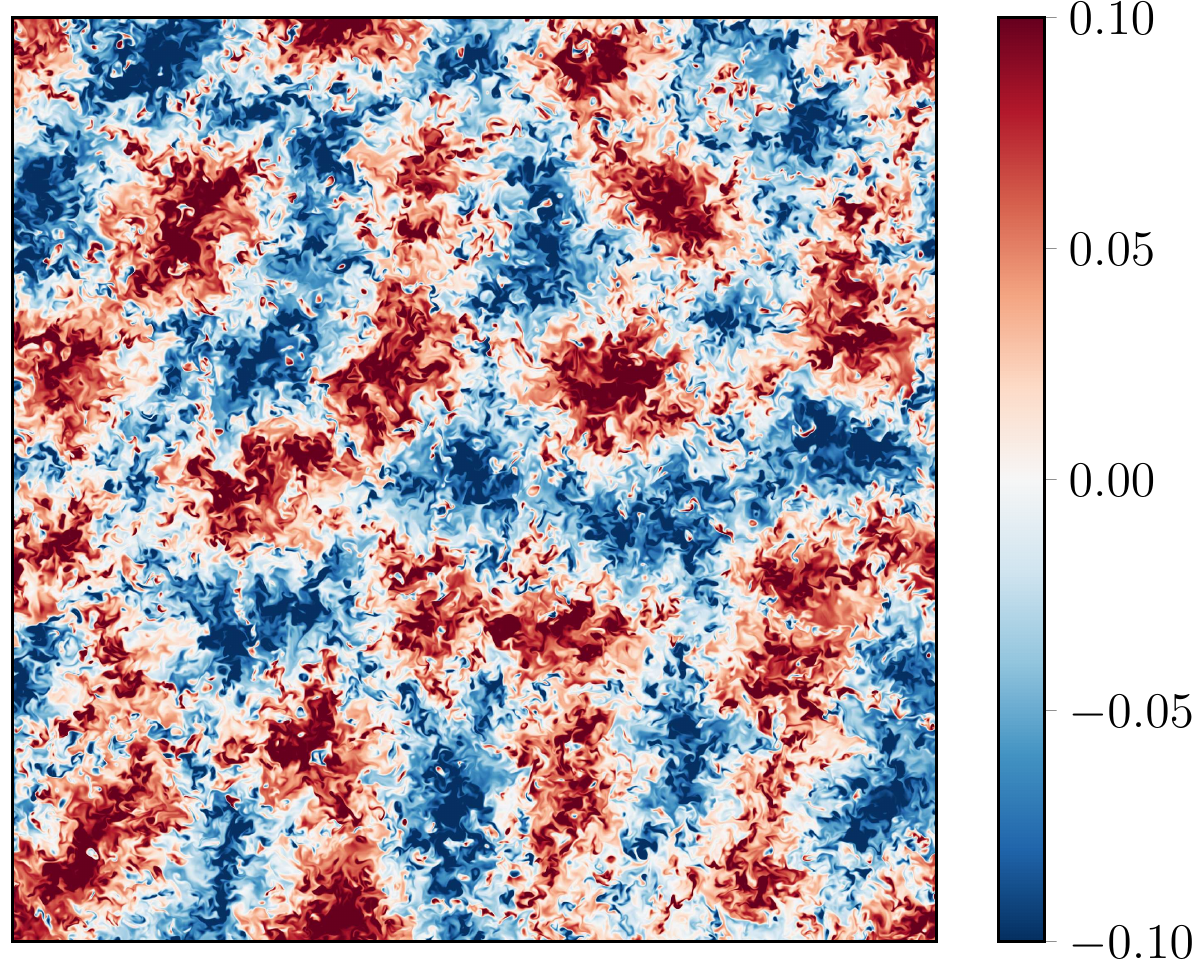}
		\caption{midplane}
	\end{subfigure}
	\begin{subfigure}{0.475\linewidth}
		\centering
		\includegraphics[width=1\linewidth]{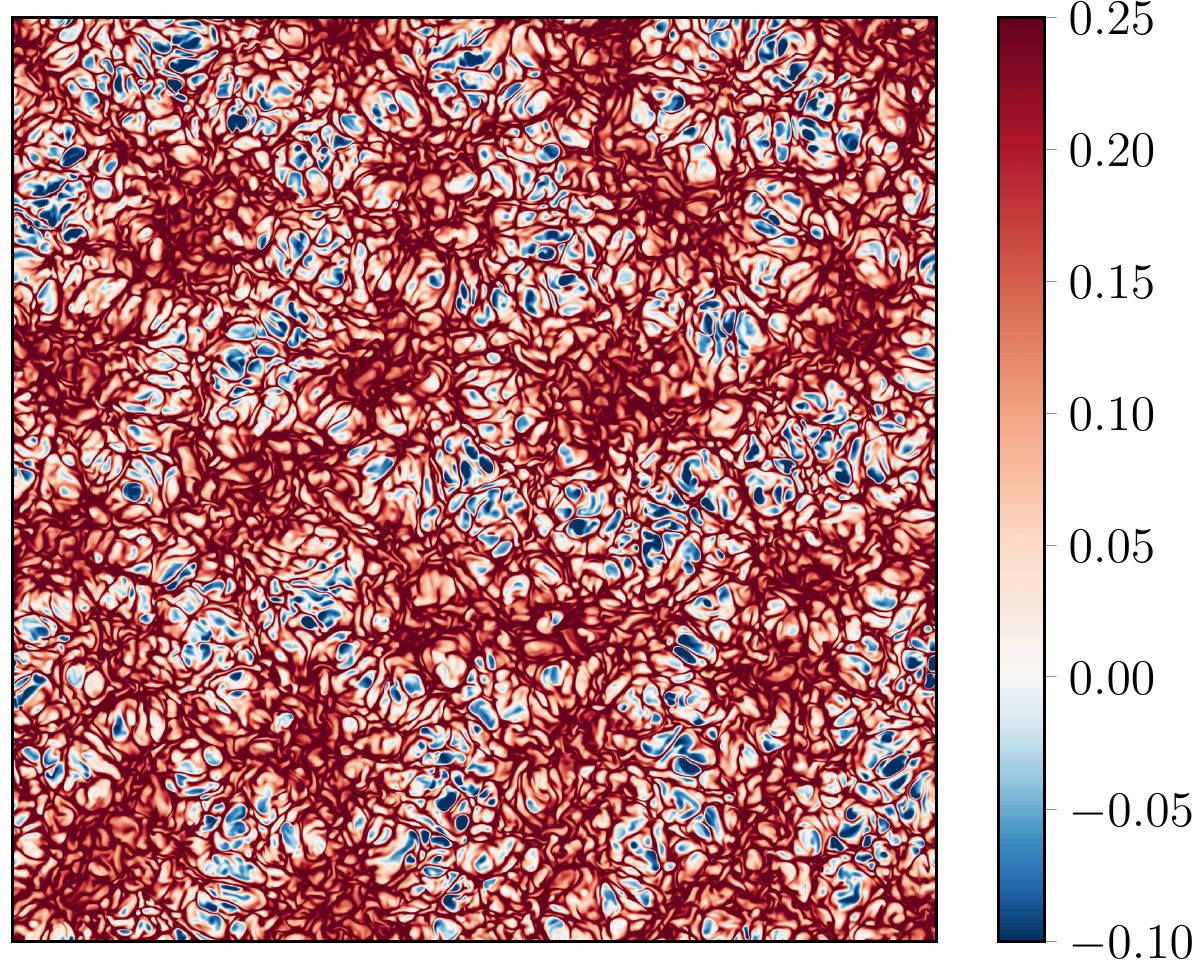}
		\caption{$z\approx 0.85 \lambda_{Nu}$}
	\end{subfigure}
	\caption{Comparison of the temperature in (a) the midplane and (b) close to the wall for $\Ra=1.07\times10^7$. A footprint of the large-scale pattern in the midplane is visible close to the bottom wall.}
	\label{fig:temp_comp}
\end{figure}
Here, we discuss the height dependence of the spectrum $E_{\theta u_z}(k,z)$ first introduced in section \ref{sec:superstructure_scale}. Because of the lack of statistical homogeneity in the vertical direction, it is not \textit{a priori} clear that there is a single characteristic large scale at all heights. 
However, it was already shown by \citet{parodi04prl,vonHardenberg08pra,pandey18natcomm,stevens18prf} and \citet{krug19arxiv} that the turbulent superstructures leave an imprint in the boundary layers. Figure \ref{fig:temp_comp} shows a comparison of the temperature field in the midplane and at boundary layer height close to the wall, which visually confirms the connection between the bulk flow and the boundary layer (see also \citet{stevens18prf}).

 \begin{figure}
 	\centering
 	\begin{subfigure}{0.475\linewidth}
 		\centering
 		\includegraphics[width=1\linewidth]{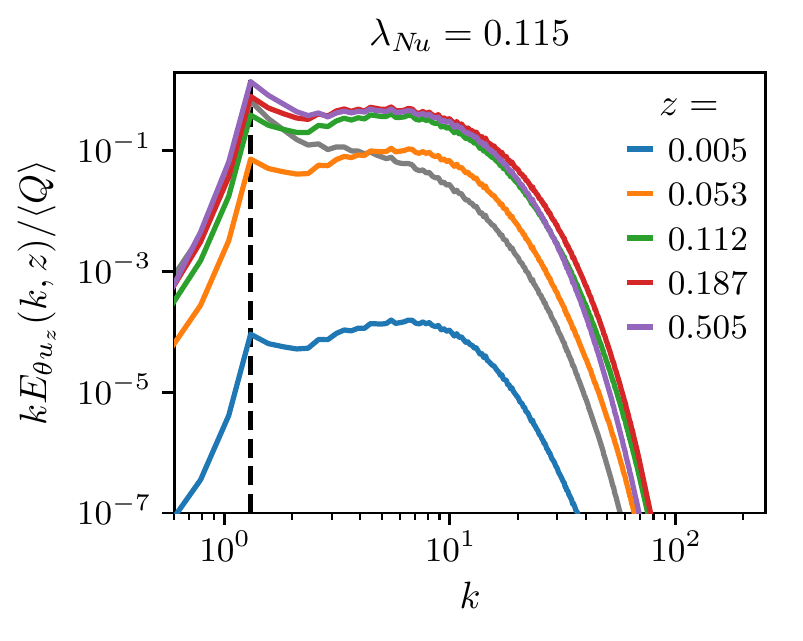}
 		\caption{$\Ra=1.02\times10^5$ }
 	\end{subfigure}
 	\begin{subfigure}{0.475\linewidth}
 		\centering
 		\includegraphics[width=1\linewidth]{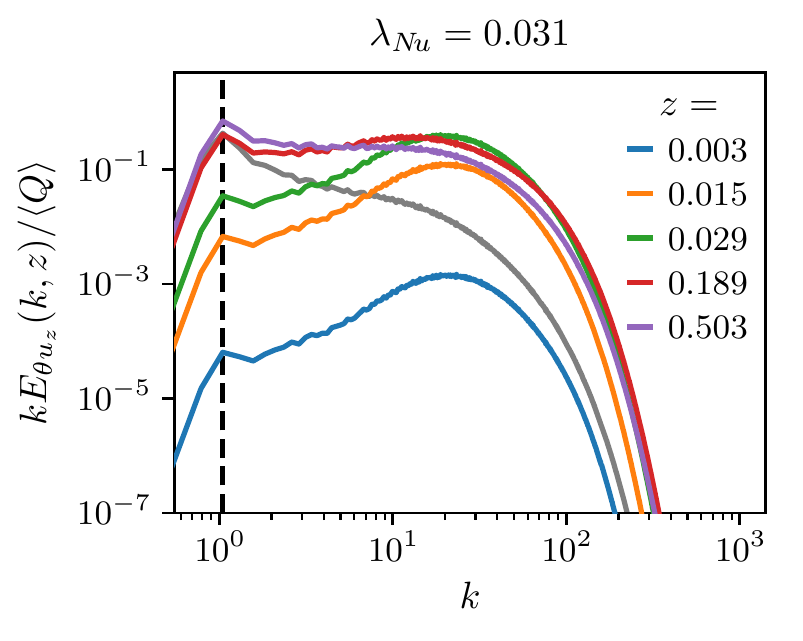}
 		\caption{$\Ra=1.07\times10^7$ }
 	\end{subfigure}
 	\caption{Pre-multiplied spectra, $kE_{\theta u_z}(k,z)$, for (a) $\Ra=1.02\times10^5$ and (b) $\Ra=1.07\times10^7$, and different heights $z$. The height-averaged spectrum is shown in dark gray. The thermal boundary layer thickness $\lambda_{Nu}$ is given for reference. A peak at the same position $k$ is present at all heights, also in the boundary layer, characterizing the size of the superstructure. However, close to the boundary layer a second maximum emerges. This is related to small-scale fluctuations. The maximum at small scales is highlighted through the presentation in pre-multiplied form.}
 	\label{fig:spec}
 \end{figure}

To verify this quantitatively, we consider the height-dependent azimuthally averaged cross-spectrum $E_{\theta u_z}(k,z)$ of the vertical velocity and temperature.
The spectrum is shown in figure \ref{fig:spec} in pre-multiplied form for two different $\Ra$ and different heights as well as height averaged.
In the midplane a single maximum is present, which characterizes the size of the superstructure. However, closer to the wall a second maximum forms \citep{kaimal76jas,mellado16blm,krug19arxiv}, which is related to the small-scale turbulent fluctuations. As expected, this maximum is more pronounced at the higher Rayleigh number. Still, we observe a local maximum at the scale of the superstructure, corresponding to the wavenumber of the maximum in the midplane. This shows that the size of the superstructure is indeed independent of height, and can also be inferred from the single peak of the height-averaged spectrum.

\section{Horizontally averaged $\Pi_l$ for varying filter scale}\label{app:Pi_l_profile}
\begin{figure}
	\centering
	\includegraphics[width=1\textwidth]{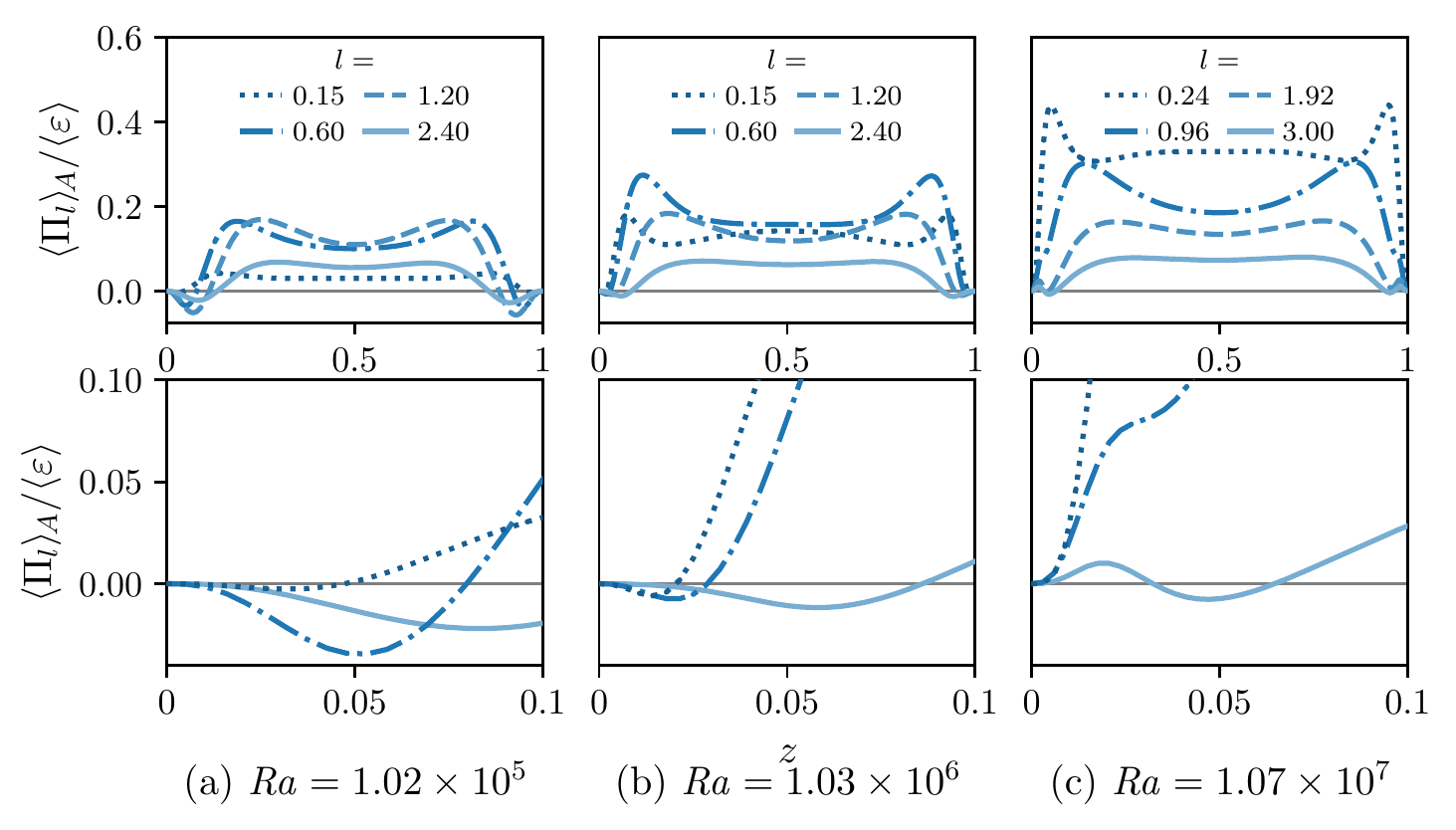} 
	\caption{Profile of the energy transfer $\avgA{\Pi_l}$ for three different $\Ra$ and different filter width $l$. The largest filter scale corresponds to the scale of the superstructures $l_s$. The bottom row shows a zoom into the region close to the wall.}
	\label{fig:Pi_l_A}
\end{figure}
Here we discuss the scale dependence of $\Pi_l$ and the corresponding profiles $\avgA{\Pi_l}$. The profiles of the energy transfer term $\Pi_l$ are strongly scale dependent, as can be expected.
Figure \ref{fig:Pi_l_A} shows the horizontally averaged energy transfer profiles for different filter widths and different $\Ra$. For $\Ra=1.02\times10^5$ and $\Ra=1.03\times10^6$, the inverse transfer layer grows in size and magnitude with increasing filter width. For $\Ra=1.07\times10^7$, the inverse energy transfer close to the wall only occurs for large filter widths. For small filter widths, the profiles are consistent with the ones for $\Ra=10^7$ reported by \citet{togni17pitvii,togni19jfm}, who obtained the height-dependent budgets for small filter width ($l<0.25$) and smaller aspect ratio ($\Gamma=8$) with a spectral cutoff filter in the horizontal directions. They found that the energy transfer acts as a dissipation for all heights throughout the layer, consistent with our findings at small filter width. 
The inverse transfer at large scales reported here indicates the need for different modelling approaches for the large-scales dynamics compared to the one at smaller scales.

\begin{figure}
	\centering
	\begin{subfigure}{1\linewidth}
		\centering
		\includegraphics[width=0.75\linewidth]{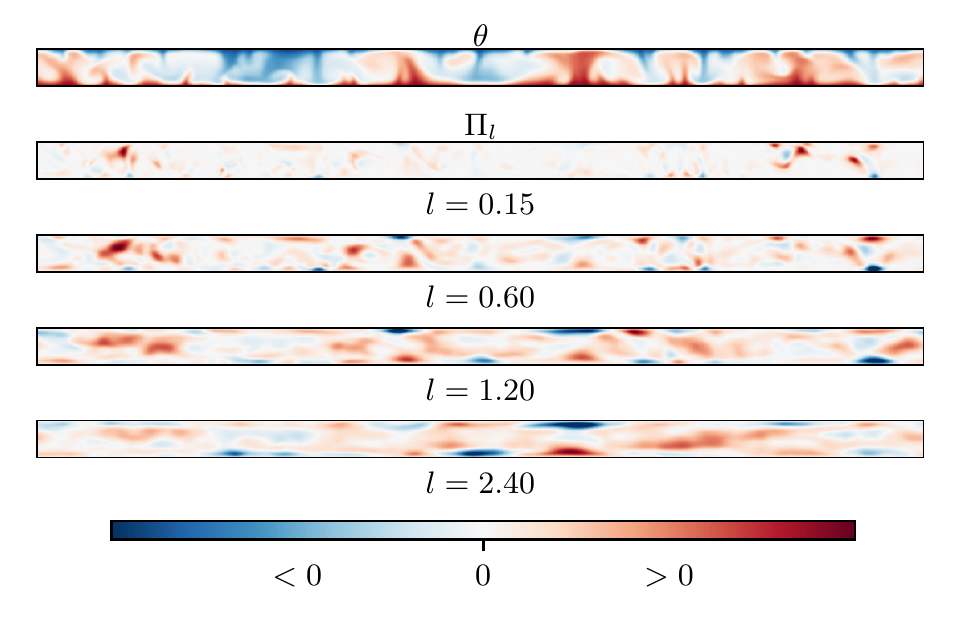}
		\caption{$\Ra=1.02\times10^5$}
		\label{fig:pi_l_local_a}
	\end{subfigure}
	\begin{subfigure}{1\linewidth}
		\centering
		\includegraphics[width=0.75\linewidth]{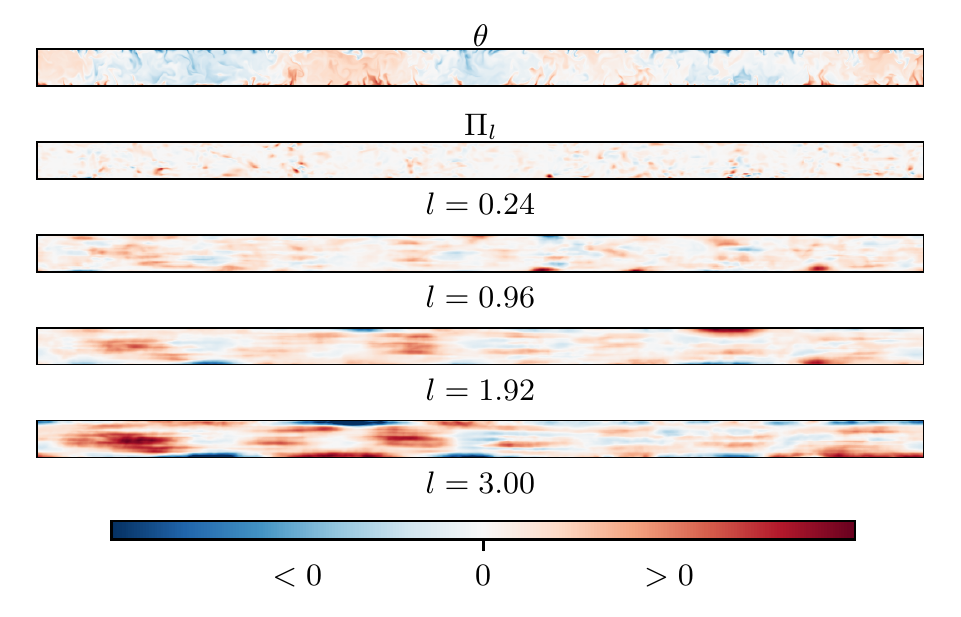}
		\caption{$\Ra=1.07\times10^7$}
		\label{fig:pi_l_local_b}
	\end{subfigure}
	\caption{Comparison of vertical cuts through the temperature fields $\theta$ and the energy transfer field $\Pi_l$ (normalized to unit maximum amplitude) for different filter widths and two $\Ra$.}
	\label{fig:pi_l_local}
\end{figure}

In section \ref{sec:TransferAndPlumes}, we related the direction of the energy transfer to the plume dynamics. Here, we discuss the inverse transfer layer at different filter widths. In figure \ref{fig:pi_l_local_a}, a vertical cut through $\Pi_l$ is shown for different filter widths and compared to the corresponding temperature field for $\Ra=1.02\times10^5$. Differently sized plumes extend through the whole cell and impinge on the wall. This causes an inverse transfer at small and large filter width. In contrast, for $\Ra=1.07\times10^7$ in figure \ref{fig:pi_l_local_b}, it can be seen that only few isolated small-scale plumes extend throughout the whole cell and impinge on the wall. They only cause little inverse energy transfer at small filter widths, which is why on average the direct transfer dominates and there is no inverse transfer layer. However, clustered plumes form larger-scale structures, which contribute to the inverse transfer at larger scales when they impinge on the wall. This results in an inverse transfer layer in the profiles at large filter widths. 

\clearpage
\bibliographystyle{jfm}
\bibliography{literature}

\begin{thebibliography}{64}
\expandafter\ifx\csname natexlab\endcsname\relax\def\natexlab#1{#1}\fi
\def\au#1{#1} \def\ed#1{#1} \def\yr#1{#1}\def\at#1{#1}\def\jt#1{\textit{#1}}
  \def\bt#1{#1}\def\bvol#1{\textbf{#1}} \def\vol#1{#1} \def\pg#1{#1}
  \def\publ#1{#1}\def\arxiv#1{#1}\def\org#1{#1}\def\st#1{\textit{#1}}

\bibitem[Ahlers {\em et~al.\/}(2009)Ahlers, Grossmann \& Lohse]{ahlers09rmp}
{\sc \au{Ahlers, G.}, \au{Grossmann, S.} \& \au{Lohse, D.}} \yr{2009}  \at{Heat
  transfer and large scale dynamics in turbulent {R}ayleigh-{B}\'{e}nard
  convection}.  \jt{Rev. Mod. Phys.}  \bvol{81},  \pg{503--537}.

\bibitem[Aluie \& Eyink(2009)]{aluie09pof}
{\sc \au{Aluie, H.} \& \au{Eyink, G.~L.}} \yr{2009}  \at{{Localness of energy
  cascade in hydrodynamic turbulence. II. Sharp spectral filter}}.  \jt{Phys.
  Fluids}  \bvol{21}~(11),  \pg{115108}.

\bibitem[Atkinson \& Zhang(1996)]{atkinson96rog}
{\sc \au{Atkinson, B.~W.} \& \au{Zhang, J.~Wu}} \yr{1996}  \at{Mesoscale
  shallow convection in the atmosphere}.  \jt{Rev. Geophys.}  \bvol{34}~(4),
  \pg{403--431}.

\bibitem[Ballouz \& Ouellette(2018)]{ballouz18jfm}
{\sc \au{Ballouz, J.~G.} \& \au{Ouellette, N.~T.}} \yr{2018}  \at{Tensor
  geometry in the turbulent cascade}.  \jt{J. Fluid Mech.}  \bvol{835},
  \pg{1048--1064}.

\bibitem[Bauer {\em et~al.\/}(2019)Bauer, von Kameke \& Wagner]{bauer19prf}
{\sc \au{Bauer, C.}, \au{von Kameke, A.} \& \au{Wagner, C.}} \yr{2019}
  \at{{Kinetic energy budget of the largest scales in turbulent pipe flow}}.
  \jt{Phys. Rev. Fluids}  \bvol{4}~(6),  \pg{064607}.

\bibitem[Bodenschatz {\em et~al.\/}(2000)Bodenschatz, Pesch \&
  Ahlers]{bodenschatz00arofm}
{\sc \au{Bodenschatz, E.}, \au{Pesch, W.} \& \au{Ahlers, G.}} \yr{2000}
  \at{Recent developments in {R}ayleigh-{B}\'{e}nard convection}.  \jt{Annu.
  Rev. Fluid Mech.}  \bvol{32}~(1),  \pg{709--778}.

\bibitem[Buzzicotti {\em et~al.\/}(2018)Buzzicotti, Linkmann, Aluie, Biferale,
  Brasseur \& Meneveau]{buzzicotti18jot}
{\sc \au{Buzzicotti, M.}, \au{Linkmann, M.}, \au{Aluie, H.}, \au{Biferale, L.},
  \au{Brasseur, J.} \& \au{Meneveau, C.}} \yr{2018}  \at{{Effect of filter type
  on the statistics of energy transfer between resolved and subfilter scales
  from a-priori analysis of direct numerical simulations of isotropic
  turbulence}}.  \jt{J. Turbul.}  \bvol{19}~(2),  \pg{167--197}.

\bibitem[Chill{\`{a}} \& Schumacher(2012)]{chilla12epje}
{\sc \au{Chill{\`{a}}, F.} \& \au{Schumacher, J.}} \yr{2012}  \at{{New
  perspectives in turbulent Rayleigh-B\'{e}nard convection}}.  \jt{Eur. Phys.
  J. E}  \bvol{35}~(7),  \pg{58}.

\bibitem[Ching(2014)]{ching14}
{\sc \au{Ching, E.~S.C.}} \yr{2014} {\em {Statistics and Scaling in Turbulent
  Rayleigh-B{\'{e}}nard Convection}\/}, 1st edn.  \publ{Springer Singapore}.

\bibitem[Cimarelli \& {De Angelis}(2011)]{cimarelli11jfm}
{\sc \au{Cimarelli, A.} \& \au{{De Angelis}, E.}} \yr{2011}  \at{Analysis of
  the {K}olmogorov equation for filtered wall-turbulent flows}.  \jt{J. Fluid
  Mech.}  \bvol{676},  \pg{376--395}.

\bibitem[Cimarelli \& {De Angelis}(2012)]{cimarelli12pof}
{\sc \au{Cimarelli, A.} \& \au{{De Angelis}, E.}} \yr{2012}  \at{{Anisotropic
  dynamics and sub-grid energy transfer in wall-turbulence}}.  \jt{Phys.
  Fluids}  \bvol{24}~(1),  \pg{015102}.

\bibitem[Cimarelli {\em et~al.\/}(2015)Cimarelli, {De Angelis}, Schlatter,
  Brethouwer, Talamelli \& Casciola]{cimarelli15jfm}
{\sc \au{Cimarelli, A.}, \au{{De Angelis}, E.}, \au{Schlatter, P.},
  \au{Brethouwer, G.}, \au{Talamelli, A.} \& \au{Casciola, C.~M.}} \yr{2015}
  \at{{Sources and fluxes of scale energy in the overlap layer of wall
  turbulence}}.  \jt{J. Fluid Mech.}  \bvol{771},  \pg{407--423}.

\bibitem[Domaradzki {\em et~al.\/}(1994)Domaradzki, Liu, H{\"{a}}rtel \&
  Kleiser]{domaradzki94pof}
{\sc \au{Domaradzki, J.~A.}, \au{Liu, W.}, \au{H{\"{a}}rtel, C.} \&
  \au{Kleiser, L.}} \yr{1994}  \at{{Energy transfer in numerically simulated
  wall-bounded turbulent flows}}.  \jt{Phys. Fluids}  \bvol{6}~(4),
  \pg{1583--1599}.

\bibitem[Elperin {\em et~al.\/}(2006{\natexlab{{\em a\/}}})Elperin, Golubev,
  Kleeorin \& Rogachevskii]{elperin06pof}
{\sc \au{Elperin, T.}, \au{Golubev, I.}, \au{Kleeorin, N.} \& \au{Rogachevskii,
  I.}} \yr{2006{\natexlab{{\em a\/}}}}  \at{{Large-scale instabilities in a
  nonrotating turbulent convection}}.  \jt{Phys. Fluids}  \bvol{18}~(12),
  \pg{126601}.

\bibitem[Elperin {\em et~al.\/}(2002)Elperin, Kleeorin, Rogachevskii \&
  Zilitinkevich]{elperin02pre}
{\sc \au{Elperin, T.}, \au{Kleeorin, N.}, \au{Rogachevskii, I.} \&
  \au{Zilitinkevich, S.}} \yr{2002}  \at{{Formation of large-scale
  semiorganized structures in turbulent convection}}.  \jt{Phys. Rev. E}
  \bvol{66}~(6),  \pg{066305}.

\bibitem[Elperin {\em et~al.\/}(2006{\natexlab{{\em b\/}}})Elperin, Kleeorin,
  Rogachevskii \& Zilitinkevich]{elperin06blm}
{\sc \au{Elperin, T.}, \au{Kleeorin, N.}, \au{Rogachevskii, I.} \&
  \au{Zilitinkevich, S.~S.}} \yr{2006{\natexlab{{\em b\/}}}}  \at{Tangling
  turbulence and semi-organized structures in convective boundary layers}.
  \jt{Bound.-Layer Meteorol.}  \bvol{119}~(3),  \pg{449--472}.

\bibitem[Emran \& Schumacher(2015)]{emran15jfm}
{\sc \au{Emran, M.~S.} \& \au{Schumacher, J.}} \yr{2015}  \at{{Large-scale mean
  patterns in turbulent convection}}.  \jt{J. Fluid Mech.}  \bvol{776},
  \pg{96--108}.

\bibitem[Eyink(1995)]{eyink95josp}
{\sc \au{Eyink, G.~L.}} \yr{1995}  \at{{Local energy flux and the refined
  similarity hypothesis}}.  \jt{J. Stat. Phys.}  \bvol{78}~(1),  \pg{335--351}.

\bibitem[Eyink(2007)]{eyink07coursenotes}
{\sc \au{Eyink, G.~L.}} \yr{2007} {Turbulence Theory, course notes, The Johns
  Hopkins University, 2007-2008}. Available at
  \url{http://www.ams.jhu.edu/~eyink/Turbulence/notes.html}.

\bibitem[Eyink \& Aluie(2009)]{eyink09pof}
{\sc \au{Eyink, G.~L.} \& \au{Aluie, H.}} \yr{2009}  \at{{Localness of energy
  cascade in hydrodynamic turbulence. I. Smooth coarse graining}}.  \jt{Phys.
  Fluids}  \bvol{21}~(11),  \pg{115107}.

\bibitem[Faranda {\em et~al.\/}(2018)Faranda, Lembo, Iyer, Kuzzay, Chibbaro,
  Daviaud \& Dubrulle]{faranda18jas}
{\sc \au{Faranda, D.}, \au{Lembo, V.}, \au{Iyer, M.}, \au{Kuzzay, D.},
  \au{Chibbaro, S.}, \au{Daviaud, F.} \& \au{Dubrulle, B.}} \yr{2018}
  \at{Computation and characterization of local subfilter-scale energy
  transfers in atmospheric flows}.  \jt{J. Atmos. Sci.}  \bvol{75}~(7),
  \pg{2175--2186}.

\bibitem[Fodor {\em et~al.\/}(2019)Fodor, Mellado \& Wilczek]{fodor19blm}
{\sc \au{Fodor, K.}, \au{Mellado, J.~P.} \& \au{Wilczek, M.}} \yr{2019}  \at{On
  the role of large-scale updrafts and downdrafts in deviations from
  {M}onin-{O}bukhov similarity theory in free convection}.  \jt{Boundary-Layer
  Meteorol.}  \bvol{172}~(3),  \pg{371--396}.

\bibitem[Fonda {\em et~al.\/}(2019)Fonda, Pandey, Schumacher \&
  Sreenivasan]{fonda19pnas}
{\sc \au{Fonda, E.}, \au{Pandey, A.}, \au{Schumacher, J.} \& \au{Sreenivasan,
  K.~R.}} \yr{2019}  \at{{Deep learning in turbulent convection networks}}.
  \jt{Proc. Natl. Acad. Sci.}  \bvol{116}~(18),  \pg{8667--8672}.

\bibitem[Germano(1992)]{germano92jfm}
{\sc \au{Germano, M.}} \yr{1992}  \at{{Turbulence: the filtering approach}}.
  \jt{J. Fluid Mech.}  \bvol{238},  \pg{325--336}.

\bibitem[Getling(1998)]{getling98}
{\sc \au{Getling, A.~V.}} \yr{1998} {\em {Rayleigh-B{\'{e}}nard Convection}\/},
   \st{Advanced Series in Nonlinear Dynamics},  \vol{vol.~11}.  \publ{World
  Scientific}.

\bibitem[Grossmann \& Lohse(2004)]{grossmann04pof}
{\sc \au{Grossmann, S.} \& \au{Lohse, D.}} \yr{2004}  \at{{Fluctuations in
  turbulent Rayleigh-B{\'{e}}nard convection: The role of plumes}}.  \jt{Phys.
  Fluids}  \bvol{16}~(12),  \pg{4462--4472}.

\bibitem[von Hardenberg {\em et~al.\/}(2008)von Hardenberg, Parodi, Passoni,
  Provenzale \& Spiegel]{vonHardenberg08pra}
{\sc \au{von Hardenberg, J.}, \au{Parodi, A.}, \au{Passoni, G.},
  \au{Provenzale, A.} \& \au{Spiegel, E.~A.}} \yr{2008}  \at{Large-scale
  patterns in {R}ayleigh-{B}\'{e}nard convection}.  \jt{Phys. Lett. A}
  \bvol{372}~(13),  \pg{2223 -- 2229}.

\bibitem[Hartlep {\em et~al.\/}(2003)Hartlep, Tilgner \& Busse]{hartlep03prl}
{\sc \au{Hartlep, T.}, \au{Tilgner, A.} \& \au{Busse, F.~H.}} \yr{2003}
  \at{Large scale structures in {R}ayleigh-{B}\'{e}nard convection at high
  rayleigh numbers}.  \jt{Phys. Rev. Lett.}  \bvol{91}~(6),  \pg{064501}.

\bibitem[Hartlep {\em et~al.\/}(2005)Hartlep, Tilgner \& Busse]{hartlep05jfm}
{\sc \au{Hartlep, T.}, \au{Tilgner, A.} \& \au{Busse, F.~H.}} \yr{2005}
  \at{{Transition to turbulent convection in a fluid layer heated from below at
  moderate aspect ratio}}.  \jt{J. Fluid Mech.}  \bvol{544},  \pg{309--322}.

\bibitem[Ibbeken {\em et~al.\/}(2019)Ibbeken, Green \& Wilczek]{ibbeken19prl}
{\sc \au{Ibbeken, G.}, \au{Green, G.} \& \au{Wilczek, M.}} \yr{2019}
  \at{Large-scale pattern formation in the presence of small-scale random
  advection}.  \jt{Phys. Rev. Lett.}  \bvol{123},  \pg{114501}.

\bibitem[Kaimal {\em et~al.\/}(1976)Kaimal, Wyngaard, Haugen, Cot{\'{e}},
  Izumi, Caughey \& Readings]{kaimal76jas}
{\sc \au{Kaimal, J.~C.}, \au{Wyngaard, J.~C.}, \au{Haugen, D.~A.},
  \au{Cot{\'{e}}, O.~R.}, \au{Izumi, Y.}, \au{Caughey, S.~J.} \& \au{Readings,
  C.~J.}} \yr{1976}  \at{Turbulence structure in the convective boundary
  layer}.  \jt{J. Atmos. Sci.}  \bvol{33}~(11),  \pg{2152--2169}.

\bibitem[Kimmel \& Domaradzki(2000)]{kimmel00pof}
{\sc \au{Kimmel, S.~J.} \& \au{Domaradzki, J.~A.}} \yr{2000}  \at{{Large eddy
  simulations of Rayleigh-B{\'{e}}nard convection using subgrid scale
  estimation model}}.  \jt{Phys. Fluids}  \bvol{12}~(1),  \pg{169--184}.

\bibitem[Krug {\em et~al.\/}(2019)Krug, Lohse \& Stevens]{krug19arxiv}
{\sc \au{Krug, D.}, \au{Lohse, D.} \& \au{Stevens, R. J. A.~M.}} \yr{2019}
  \at{{Coherence of temperature and velocity superstructures in turbulent
  Rayleigh-B\'{e}nard flow}}.  \jt{J. Fluid Mech. \textnormal{(in press)}} ,
  \arxiv{arXiv: 1908.10073}.

\bibitem[Lohse \& Xia(2010)]{lohse10arofm}
{\sc \au{Lohse, D.} \& \au{Xia, K.-Q.}} \yr{2010}  \at{Small-scale properties
  of turbulent {R}ayleigh-{B}\'{e}nard convection}.  \jt{Annu. Rev. Fluid
  Mech.}  \bvol{42}~(1),  \pg{335--364}.

\bibitem[Lomax {\em et~al.\/}(2001)Lomax, Pulliam \& Zingg]{lomax01}
{\sc \au{Lomax, H.}, \au{Pulliam, T.~H.} \& \au{Zingg, D.~W.}} \yr{2001} {\em
  {Fundamentals of Computational Fluid Dynamics}\/}, 1st edn. {\em Scientific
  Computation\/} .  \publ{Springer Berlin Heidelberg}.

\bibitem[L{\"{u}}lff {\em et~al.\/}(2011)L{\"{u}}lff, Wilczek \&
  Friedrich]{lulff11njp}
{\sc \au{L{\"{u}}lff, J.}, \au{Wilczek, M.} \& \au{Friedrich, R.}} \yr{2011}
  \at{{Temperature statistics in turbulent Rayleigh-B{\'{e}}nard convection}}.
  \jt{New J. Phys.}  \bvol{13}~(1),  \pg{015002}.

\bibitem[L{\"{u}}lff {\em et~al.\/}(2015)L{\"{u}}lff, Wilczek, Stevens,
  Friedrich \& Lohse]{lulff15jfm}
{\sc \au{L{\"{u}}lff, J.}, \au{Wilczek, M.}, \au{Stevens, R. J. A.~M.},
  \au{Friedrich, R.} \& \au{Lohse, D.}} \yr{2015}  \at{{Turbulent
  Rayleigh-B{\'{e}}nard convection described by projected dynamics in phase
  space}}.  \jt{J. Fluid Mech.}  \bvol{781},  \pg{276--297}.

\bibitem[Manneville(1990)]{manneville90}
{\sc \au{Manneville, P.}} \yr{1990} {\em {Dissipative Structures and Weak
  Turbulence}\/}.  \publ{Elsevier}.

\bibitem[Marati {\em et~al.\/}(2004)Marati, Casciola \& Piva]{marati04jfm}
{\sc \au{Marati, N.}, \au{Casciola, C.~M.} \& \au{Piva, R.}} \yr{2004}
  \at{{Energy cascade and spatial fluxes in wall turbulence}}.  \jt{J. Fluid
  Mech.}  \bvol{521},  \pg{191--215}.

\bibitem[Mellado(2012)]{mellado12jfm}
{\sc \au{Mellado, J.~P.}} \yr{2012}  \at{{Direct numerical simulation of free
  convection over a heated plate}}.  \jt{J. Fluid Mech.}  \bvol{712},
  \pg{418--450}.

\bibitem[Mellado \& Ansorge(2012)]{mellado12zamm}
{\sc \au{Mellado, J.~P.} \& \au{Ansorge, C.}} \yr{2012}  \at{Factorization of
  the {F}ourier transform of the pressure-{P}oisson equation using finite
  differences in colocated grids}.  \jt{Z. Angew. Math. Mech.}  \bvol{92}~(5),
  \pg{380--392}.

\bibitem[Mellado {\em et~al.\/}(2016)Mellado, van Heerwaarden \&
  Garcia]{mellado16blm}
{\sc \au{Mellado, J.~P.}, \au{van Heerwaarden, C.~C.} \& \au{Garcia, J.~R.}}
  \yr{2016}  \at{Near-surface effects of free atmosphere stratification in free
  convection}.  \jt{Boundary-Layer Meteorol.}  \bvol{159}~(1),  \pg{69--95}.

\bibitem[Morris {\em et~al.\/}(1993)Morris, Bodenschatz, Cannell \&
  Ahlers]{morris93prl}
{\sc \au{Morris, S.~W.}, \au{Bodenschatz, E.}, \au{Cannell, D.~S.} \&
  \au{Ahlers, G.}} \yr{1993}  \at{{Spiral defect chaos in large aspect ratio
  Rayleigh-B{\'{e}}nard convection}}.  \jt{Phys. Rev. Lett.}  \bvol{71}~(13),
  \pg{2026--2029}.

\bibitem[Nordlund {\em et~al.\/}(2009)Nordlund, Stein \&
  Asplund]{nordlund09lrisp}
{\sc \au{Nordlund, {\AA}.}, \au{Stein, R.~F.} \& \au{Asplund, M.}} \yr{2009}
  \at{Solar surface convection}.  \jt{Living Rev. Sol. Phys.}  \bvol{6}~(1),
  \pg{2}.

\bibitem[Pandey {\em et~al.\/}(2018)Pandey, Scheel \&
  Schumacher]{pandey18natcomm}
{\sc \au{Pandey, A.}, \au{Scheel, J.~D.} \& \au{Schumacher, J.}} \yr{2018}
  \at{Turbulent superstructures in {R}ayleigh-{B}\'{e}nard convection}.
  \jt{Nat. Commun.}  \bvol{9}~(1),  \pg{2118}.

\bibitem[Parodi {\em et~al.\/}(2004)Parodi, von Hardenberg, Passoni, Provenzale
  \& Spiegel]{parodi04prl}
{\sc \au{Parodi, A.}, \au{von Hardenberg, J.}, \au{Passoni, G.},
  \au{Provenzale, A.} \& \au{Spiegel, E.~A.}} \yr{2004}  \at{Clustering of
  plumes in turbulent convection}.  \jt{Phys. Rev. Lett.}  \bvol{92},
  \pg{194503}.

\bibitem[Petschel {\em et~al.\/}(2013)Petschel, Stellmach, Wilczek, L{\"{u}}lff
  \& Hansen]{petschel13prl}
{\sc \au{Petschel, K.}, \au{Stellmach, S.}, \au{Wilczek, M.}, \au{L{\"{u}}lff,
  J.} \& \au{Hansen, U.}} \yr{2013}  \at{Dissipation layers in
  {R}ayleigh-{B}\'{e}nard convection: {A} unifying view}.  \jt{Phys. Rev.
  Lett.}  \bvol{110}~(11),  \pg{114502}.

\bibitem[Petschel {\em et~al.\/}(2015)Petschel, Stellmach, Wilczek, L{\"{u}}lff
  \& Hansen]{petschel15jfm}
{\sc \au{Petschel, K.}, \au{Stellmach, S.}, \au{Wilczek, M.}, \au{L{\"{u}}lff,
  J.} \& \au{Hansen, U.}} \yr{2015}  \at{Kinetic energy transport in
  {R}ayleigh-{B}\'{e}nard convection}.  \jt{J. Fluid Mech.}  \bvol{773},
  \pg{395--417}.

\bibitem[Pope(2000)]{pope2000}
{\sc \au{Pope, S.~B.}} \yr{2000} {\em {Turbulent Flows}\/}.  \publ{Cambridge
  University Press}.

\bibitem[Sagaut(2006)]{sagaut06}
{\sc \au{Sagaut, P.}} \yr{2006} {\em {Large Eddy Simulation for Incompressible
  Flows}\/}, 3rd edn.  \publ{Springer}.

\bibitem[Scheel \& Schumacher(2014)]{scheel14jfm}
{\sc \au{Scheel, J.~D.} \& \au{Schumacher, J.}} \yr{2014}  \at{{Local boundary
  layer scales in turbulent Rayleigh-B{\'{e}}nard convection}}.  \jt{J. Fluid
  Mech.}  \bvol{758},  \pg{344--373}.

\bibitem[Schumacher {\em et~al.\/}(2018)Schumacher, Pandey, Yakhot \&
  Sreenivasan]{schumacher18pre}
{\sc \au{Schumacher, J.}, \au{Pandey, A.}, \au{Yakhot, V.} \& \au{Sreenivasan,
  K.~R.}} \yr{2018}  \at{{Transition to turbulence scaling in
  Rayleigh-B{\'{e}}nard convection}}.  \jt{Phys. Rev. E}  \bvol{98}~(3),
  \pg{033120}.

\bibitem[Shishkina {\em et~al.\/}(2010)Shishkina, Stevens, Grossmann \&
  Lohse]{shishkina10njp}
{\sc \au{Shishkina, O.}, \au{Stevens, R. J. A.~M.}, \au{Grossmann, S.} \&
  \au{Lohse, D.}} \yr{2010}  \at{Boundary layer structure in turbulent thermal
  convection and its consequences for the required numerical resolution}.
  \jt{New J. Phys.}  \bvol{12}~(7),  \pg{075022}.

\bibitem[Shishkina \& Wagner(2006)]{shishkina06jfm}
{\sc \au{Shishkina, O.} \& \au{Wagner, C.}} \yr{2006}  \at{Analysis of thermal
  dissipation rates in turbulent {R}ayleigh-{B}\'{e}nard convection}.  \jt{J.
  Fluid Mech.}  \bvol{546},  \pg{51--60}.

\bibitem[Shraiman \& Siggia(1990)]{shraiman90pra}
{\sc \au{Shraiman, B.~I.} \& \au{Siggia, E.~D.}} \yr{1990}  \at{{Heat transport
  in high-Rayleigh-number convection}}.  \jt{Phys. Rev. A}  \bvol{42}~(6),
  \pg{3650--3653}.

\bibitem[Siggia(1994)]{siggia94arof}
{\sc \au{Siggia, E.~D.}} \yr{1994}  \at{High {R}ayleigh number convection}.
  \jt{Annu. Rev. Fluid Mech.}  \bvol{26}~(1),  \pg{137--168}.

\bibitem[Stevens {\em et~al.\/}(2018)Stevens, Blass, Zhu, Verzicco \&
  Lohse]{stevens18prf}
{\sc \au{Stevens, R. J. A.~M.}, \au{Blass, A.}, \au{Zhu, X.}, \au{Verzicco, R.}
  \& \au{Lohse, D.}} \yr{2018}  \at{Turbulent thermal superstructures in
  {R}ayleigh-{B}\'{e}nard convection}.  \jt{Phys. Rev. Fluids}  \bvol{3}~(4),
  \pg{041501}.

\bibitem[Togni {\em et~al.\/}(2015)Togni, Cimarelli \& De~Angelis]{togni15jfm}
{\sc \au{Togni, R.}, \au{Cimarelli, A.} \& \au{De~Angelis, E.}} \yr{2015}
  \at{Physical and scale-by-scale analysis of {R}ayleigh-{B}\'{e}nard
  convection}.  \jt{J. Fluid Mech.}  \bvol{782},  \pg{380--404}.

\bibitem[Togni {\em et~al.\/}(2017)Togni, Cimarelli \& {De
  Angelis}]{togni17pitvii}
{\sc \au{Togni, R.}, \au{Cimarelli, A.} \& \au{{De Angelis}, E.}} \yr{2017}
  Towards an improved subgrid-scale model for thermally driven flows.  \bt{In
  {\em Progress in Turbulence VII\/} (ed. \ed{Ramis {\"{O}}rl{\"{u}},
  Alessandro Talamelli, Martin Oberlack \& Joachim Peinke})},  \pg{pp.
  141--145}.  \publ{Cham: Springer International Publishing}.

\bibitem[Togni {\em et~al.\/}(2019)Togni, Cimarelli \& {De
  Angelis}]{togni19jfm}
{\sc \au{Togni, R.}, \au{Cimarelli, A.} \& \au{{De Angelis}, E.}} \yr{2019}
  \at{{Resolved and subgrid dynamics of Rayleigh-B{\'{e}}nard convection}}.
  \jt{J. Fluid Mech.}  \bvol{867},  \pg{906--933}.

\bibitem[Valori {\em et~al.\/}(2020)Valori, Innocenti, Dubrulle \&
  Chibbaro]{valori20jfm}
{\sc \au{Valori, V.}, \au{Innocenti, A.}, \au{Dubrulle, B.} \& \au{Chibbaro,
  S.}} \yr{2020}  \at{Weak formulation and scaling properties of energy fluxes
  in three-dimensional numerical turbulent rayleigh-b\'enard convection}.
  \jt{J. Fluid Mech.}  \bvol{885},  \pg{A14}.

\bibitem[Verma(2018)]{verma18}
{\sc \au{Verma, M.~K.}} \yr{2018} {\em {Physics of Buoyant Flows}\/}.
  \publ{World Scientific}.

\bibitem[Verma {\em et~al.\/}(2017)Verma, Kumar \& Pandey]{verma17njp}
{\sc \au{Verma, M.~K.}, \au{Kumar, A.} \& \au{Pandey, A.}} \yr{2017}
  \at{Phenomenology of buoyancy-driven turbulence: recent results}.  \jt{New J.
  Phys.}  \bvol{19}~(2),  \pg{025012}.

\bibitem[Verzicco \& Camussi(2003)]{verzicco03jfm}
{\sc \au{Verzicco, R.} \& \au{Camussi, R.}} \yr{2003}  \at{Numerical
  experiments on strongly turbulent thermal convection in a slender cylindrical
  cell}.  \jt{J. Fluid Mech.}  \bvol{477},  \pg{19--49}.

\end{thebibliography}
\end{document}